\documentclass[aps,twocolumn,prd,superscriptaddress,nofootinbib,showpacs,letterpaper,eqsecnum]{revtex4}

\usepackage{amsmath,amssymb}
\usepackage[dvips]{graphicx}
\usepackage{longtable}
\usepackage[usenames,dvipsnames]{xcolor}
\usepackage[normalem]{ulem}
\usepackage{bm}
\usepackage{comment}
\usepackage{xcolor}
\usepackage{soul}
\usepackage{hyperref}

\begin{document}

\newcommand{\E}{\mathrm{E}}
\newcommand{\Var}{\mathrm{Var}}
\newcommand{\bra}[1]{\langle #1|}
\newcommand{\ket}[1]{|#1\rangle}
\newcommand{\braket}[2]{\langle #1|#2 \rangle}
\newcommand{\mean}[2]{\langle #1 #2 \rangle}
\newcommand{\be}{\begin{equation}}
\newcommand{\ee}{\end{equation}}	
\newcommand{\ba}{\begin{eqnarray}}
\newcommand{\ea}{\end{eqnarray}}
\newcommand{\SD}[1]{{\color{magenta}#1}}
\newcommand{\rem}[1]{{\sout{#1}}}
\newcommand{\alert}[1]{\textbf{\color{red} \uwave{#1}}}
\newcommand{\Y}[1]{\textcolor{blue}{#1}}
\newcommand{\R}[1]{\textcolor{red}{#1}}
\newcommand{\B}[1]{\textcolor{black}{#1}}
\newcommand{\C}[1]{\textcolor{cyan}{#1}}
\newcommand{\db}{\color{darkblue}}
\newcommand{\aaron}[1]{\textcolor{cyan}{#1}}
\newcommand{\fan}[1]{\textcolor{blue}{#1}}
\newcommand{\ac}[1]{\textcolor{cyan}{\sout{#1}}}
\newcommand{\intinfty}{\int_{-\infty}^{\infty}\!}
\newcommand{\Tr}{\mathop{\rm Tr}\nolimits}
\newcommand{\const}{\mathop{\rm const}\nolimits}
\newcommand{\Caltech}{\affiliation{Theoretical Astrophysics 350-17, California Institute of Technology, Pasadena, CA 91125, USA}}

\title{Quasinormal modes of nearly extremal Kerr spacetimes: spectrum bifurcation and power-law ringdown}

\author{Huan Yang} \Caltech
\author{Aaron Zimmerman} \Caltech
\author{An{\i}l Zengino\u{g}lu}\Caltech
\author{Fan Zhang} \Caltech
\author{Emanuele Berti}
\affiliation{Department of Physics and Astronomy, The University of Mississippi, University, MS 38677, USA}
\author{Yanbei Chen} \Caltech

\date{\today}

\begin{abstract}
We provide an in-depth investigation of quasinormal-mode oscillations of Kerr black holes with nearly extremal angular momenta. We first discuss in greater detail the two distinct types of quasinormal mode frequencies
presented in a recent paper \cite{Yang2012b}.
One set of modes, that we call ``zero-damping modes'', has vanishing imaginary part in the extremal limit, and exists for all corotating perturbations (i.e., modes with azimuthal index $m \geq 0$). The other set (the ``damped modes'') retains a finite decay rate even for extremal Kerr black holes, and exists only for a subset of corotating modes. As the angular momentum approaches its extremal value, the frequency spectrum bifurcates into these two distinct branches when both types of modes are present. 
We discuss the physical reason for the mode branching by developing and using a bound-state formulation for the perturbations of generic Kerr black holes. 
We also numerically explore the specific case of the fundamental $l=2$ modes, which have the greatest astrophysical interest. 
Using the results of these investigations, we compute the quasinormal mode response of a nearly extremal Kerr black hole to perturbations. We show that many superimposed overtones result in a slow power-law decay of the quasinormal ringing at early times, which later gives way to exponential decay. This exceptional early-time power-law decay implies that the ringdown phase is long-lived for black holes with large angular momentum, which could provide a promising strong source for gravitational-wave detectors.
\end{abstract}


\maketitle

\section{Introduction}

\subsection{Motivation}
Astrophysical black holes are naturally rotating, as they generally inherit nonzero angular momentum from their progenitors (e.g. stellar mass compact binary mergers, the gravitational collapse of massive stars, and merging supermassive black holes). According to the weak cosmic censorship conjecture, these black holes have angular momenta bounded by their masses: in geometrical units ($G=c=1$) $J \le M^2$, or equivalently $a\equiv J/M^2 \le 1$, where $a$ is the dimensionless spin parameter. In fact, as shown by Thorne \cite{Thorne}, thin accretion disks could in principal spin up their central black holes to $a=0.998$. More realistic numerical models of accretion suggest that the actual limit may be lower \cite{Gammie:2003qi}, but observations and theoretical arguments imply that fast-spinning black holes should be ubiquitous in the Universe \cite{Berti:2008af,Brenneman:2011wz,Reynolds:2013rva}, and therefore their study is astrophysically important.

Extremal black holes are also interesting for quantum field theory. As a Kerr black hole approaches the extremal limit $a\rightarrow 1$, the surface gravity of the event horizon asymptotes zero, and the near-horizon geometry reduces to AdS$_2\times$ S$^2$ \cite{bardeen}. These properties allow one to draw connections between the near-horizon quantum states and those of a two-dimensional conformal field theory \cite{Guica:2008mu}, and enable the counting of black hole entropy \cite{strominger}. 

There is also recent interest into the classical properties of extremal black holes. Generically, perturbations of subextremal Kerr black holes decay in time \cite{Kokkotas:1999bd,Berti:2009kk,Dafermos:2010hb}. It has been recently shown, however, that perturbations of spacetimes with extremal horizons have conserved quantities along the black hole horizon \cite{Aretakis:2012ei,Aretakis:2012bm}. As a consequence, transverse derivatives of the perturbations across the horizons blow-up. This may indicate nonlinear instability of extremal Kerr spacetimes (see \cite{Murata:2013daa} for a recent nonlinear study on the spherically symmetric case). These observations provide further motivation to study such spacetimes.

Spacetimes with nearly extremal black holes have unique features that make them an interesting class.  In this paper, we examine the quasinormal mode (QNM) spectrum and response of NEK black holes. QNMs play a prominent role in the gravitational emission of astrophysical black holes and the quantum behavior of dual field theories. Furthermore, they provide the most basic entry into the study of the stability of black hole solutions.

\subsection{Quasinormal modes of nearly extremal Kerr black holes}

QNMs of generic Kerr black holes are indexed by angular ``quantum numbers" $(l,\,m)$ and by an ``overtone number'' $n$, that sorts them by the magnitude of their imaginary part. This classification implicitly assumes that for each $(l,\,m)$, there is a {\em single} branch of QNMs, which is indexed by $n$. However, as originally mentioned by Leaver \cite{leaver2} for the $l=2,\, m=1$ mode and explicitly shown in \cite{Yang2012b} for general modes, part of the QNM branches of near-extremal Kerr black holes bifurcate\footnote{Though~\cite{Yang2012b} claims to discover this bifurcation effect of the Kerr spectrum, it was actually first found by Leaver for the single case of $l=2$, $m=1$~\cite{leaver}. Leaver's result appears to have been forgotten, and certainly never explored in a more general context.}. 
In this bifurcation, a single branch of QNMs splits into two branches: one set of modes reside near the horizon (see \cite{Yang2012b} or Sec.~\ref{sec:Bifurcation} for detailed discussion), with frequencies approaching the real axis as $a \rightarrow 1$, and we call them ``zero-damping modes'' (ZDMs); the other set of modes reside near the peak of the radial Teukolsky potential (see Sec.~\ref{sec:WKB}), with complex frequencies even in the extremal Kerr limit, and we refer to them as ``damped modes'' (DMs). 
For a given $l$, ZDMs exist for all $m \ge 0$ and DMs only exist in a smaller regime (approximately $0 \le m/(l+1/2) \le 0.74$). 
As a result, the NEK QNM spectrum naturally separates into two regions: one with only ZDMs (single-phase region) and the other with both sets of modes (double-phase region).

The ZDMs of NEK black holes were originally discovered by Detweiler \cite{detweiler} for $l=m$ modes using matched asymptotic expansion techniques (also see Sec.~\ref{sec:NearlyExtreme}), and later discussed by Sasaki and Nakamura \cite{sasaki} in terms of analytical frequency formulae. 
In~\cite{cardoso}, Cardoso discussed the range of validity of the Detweiler-Sasaki-Nakamura formula. 
However, the calculation by Detweiler, Sasaki, and Nakamura includes a mistake, and as a result the QNM frequencies they found are always complex, even if $a=1$. 
Hod \cite{hod2} first derived the correct QNM frequency formula in the near-extremal limit. In the previous paper \cite{Yang2012b}, we mentioned that the ZDMs, which are approximately described by the Hod formula, exist for all $m \ge 0$ in the near extremal limit. 
In this paper we prove this statement and obtain the error term in the Hod formula. 
This error term is generally small as we take the limit $a \rightarrow 1$, except for the cases where the corresponding modes are located near the boundary between the single-phase and the double-phase regions. 

Andersson and Glampedakis \cite{Glampedakis:2001js} applied the Detweiler-Sasaki-Nakamura formula to argue that there could be long-lived radiation in NEK spacetimes (more specifically, a $1/t$ power-law tail).
Here we reanalyze the problem using the Hod formula for ZDMs, and show that there is indeed a power-law decay in the early part of the signal, which turns into exponential decay later on, and eventually becomes a polynomial tail.

DMs are not captured by the matched asymptotic expansion method, partially because their mode frequencies do not satisfy the assumptions made in performing the matched asymptotic expansion (see Sec.~\ref{sec:NearlyExtreme}). 
One simple way to understand DMs is to consider QNMs in the WKB picture and then take the $a \rightarrow 1$ limit \cite{Yang2012a, Yang2012b}. 
For some of the modes in the NEK QNM spectrum (those whose quantum numbers approximately satisfy $m/(l+1/2) \ge 0.74$), the corresponding peak in the radial Teukolsky potential asymptotes to the horizon as $a \rightarrow 1$. 
These WKB modes are the ZDMs, with the imaginary part of the frequencies approaching zero. The rest of the WKB modes (the DMs) reside near WKB peaks which are outside of the horizon, even in the extremal angular momentum limit. 
These WKB peaks have finite height and width, so that they can only support a finite number of DMs. Numerical investigation shows that the fundamental DM frequency is well described by the WKB frequency formula, which supports our interpretation.

In order to better understand the mode bifurcation, we also introduce a technique which transforms the radial Teukolsky equation from a scattering problem into a bound-state problem. 
The modes in the original scattering picture and the new bound-state picture are in a one-to-one correspondence. 
In addition, the bound-state ``QNMs" have finite support in the radial domain, which enables us to visualize them by drawing their wave-functions (as in Fig.~\ref{fig:wavefunctionnearhorizon}). 
It is then clear that in the bound state picture, the ZDMs have support near the horizon and the DMs have support near the potential barriers which are outside of the horizon. This transformation was originally introduced by Mashhoon \cite{mashhoon} for Schwarzschild and slowly-rotating Kerr black holes in the eikonal limit, and we generalize it to generic Kerr black holes and generic QNMs.
 
 Our work sheds further light on the rich structure of the Kerr QNM spectrum. For generic Kerr-de Sitter black holes, the distribution of QNMs and statistics of frequencies are discussed in a recent paper by Dyatlov and Zworski \cite{Dyatlov:2013fua}, who find that the presence of ZDMs at high angular momenta limit their ability to bound the decay of perturbations in these spacetimes.

\subsection{Overview of the paper}
\label{sec:Overview}

The paper is organized as follows.

In Sec.~\ref{sec:WKB} we review the Teukolsky equation describing perturbations of the Kerr spacetime, the application of the WKB method to the Teukolsky equation in the eikonal approximation, and the resulting formula for the QNM frequencies, following~\cite{Yang2012a}. We also present the transformation of the Teukolsky equation into a bound-state problem. Section~\ref{sec:NearlyExtreme} reviews the theory of perturbations of Kerr in the NEK approximation, including WKB results and matched asymptotic expansions, presents the expression for the corresponding QNM frequencies, and discusses the leading-order correction to this formula, pointing out the existence of two different classes of NEK QNMs (Sec.~\ref{sec:PhaseTrans}).
In Sec.~\ref{sec:Bifurcation} we discuss the nature of these two phase regimes (Sec.~\ref{sec:NumPhase}), the branching behavior that must occur as a moderately spinning Kerr black hole transitions into the NEK regime (Sec.~\ref{sec:NumBifurcation}), and the physical intuition by which we can understand this behavior. Section~\ref{sec:NumBifurcation} also contains numerical investigations of some representative modes. We detail our second application of the NEK frequency formula in Sec.~\ref{sec:Decay}, where we return to the problem of the superposition of many weakly damped QNMs into a single coherent superposition of QNMs (a ``super-mode'') ringing with a power-law decay. The possibility of this behavior was first discussed and described in~\cite{Glampedakis:2001js,Andersson:1999wj}, but using an incorrect formula for the NEK frequencies. In Sec.~\ref{sec:DecayCalc} we show that our correct formula also gives a power-law decay for the early QNM response, eventually giving way to the exponential decay of the least-damped QNM. In order to check our approximations, in Sec.~\ref{sec:NumDecay} we present numerical results by solving the wave equation for large angular momenta, and we recover the early-time power-law behavior of the ringdown modes. In Sec.~\ref{sec:Conclusions} we discuss possible extensions of this work. Technical results and the discussion of our numerical methods are given in the Appendices.

Throughout this paper we use geometric units $G = c =1$, and we set the mass of the black hole to unity (i.e. $M=1$, in contrast with much of the literature on QNMs \cite{Kokkotas:1999bd,Berti:2009kk}, which sets $2 M =1$).

\section{Teukolsky equation: Bound-state formulation and WKB analysis}
\label{sec:WKB}
In this section we review the WKB approximation to the Teukolsky equation for computing QNMs of Kerr black holes. The WKB approximation is based on the eikonal and the geometric optics approximations, where perturbations propagate along null geodesics of the background. In this picture the QNMs are associated with bound, unstable photon orbits. 
This correspondence provides physical intuition into the damping rates of QNMs. The ZDMs are associated with nearly equatorial photon orbits, which asymptote to the horizon in the extremal limit. The damped modes (DMs) are associated with nearly polar photon orbits, which remain distinct from the horizon in the extremal limit. To prepare the discussion of these modes in the next section, we derive analytic formulae for the angular eigenvalues ${}_s A_{lm \omega}$ and the QNM frequencies $\omega_{lm}$. We derive these relations from the Teukolsky equation via the WKB method, supplemented with the additional assumption $(a \omega_R/L)^2 \ll 1$, as detailed in Sec.~\ref{EikWKB} below.

\subsection{The Teukolsky equation in Kerr spacetime}

The Kerr metric describes a two-parameter family of line elements, the parameters being the mass $M$ and specific angular momentum $a$~\cite{Kerr}. The line element in Boyer-Lindquist coordinates reads~\cite{BoyerLind}
\begin{subequations}
\label{eq:BL}
\begin{align}
ds^2  = & -\left( 1 - \frac{2 M r}{\rho^2} \right) dt^2 - \frac{4 M r a \sin^2 \theta}{\rho^2} dt d \phi + \frac{\rho^2}{\Delta} dr^2 \notag \\
&+ \rho^2  d \theta^2  + \left( r^2 + a ^2 + \frac{2 M r a^2 \sin^2 \theta}{\rho^2} \right) \sin^2 \theta d\phi^2 , \\
\rho^2  = & r^2 + a^2 \cos^2 \theta , \\
\Delta = & r^2 -2 M r +a^2 .
\end{align}
\end{subequations}
We set $M=1$, and therefore make no distinction between the specific angular momentum $a$ and the angular momentum $a M$.
Perturbations of the Kerr spacetime are most conveniently expressed in the Newman-Penrose formalism~\cite{NewmanPenrose}, where the Einstein field equations are projected onto a complex, null tetrad. Using these equations, Teukolsky~\cite{Teukolsky1} showed that scalar, electromagnetic, and gravitational perturbations are governed by a single master equation for scalar quantities ${}_s\psi$ of spin weight $s$. The master function ${}_s\psi$ corresponds to scalar perturbations for $s = 0$, electromagnetic perturbations for $s = \pm 1$, and gravitational perturbations for $s = \pm 2$. The master equation is separable when ${}_s\psi$ is expanded as
\begin{align}
\label{eq:MasterSep}
{}_s\psi = \sum_{l,m}\int d \omega \, {}_s R_{lm\omega} (r) \, {}_s S_{lm \omega} (\theta) e^{-i \omega t}e^{im \phi} \,.
\end{align}
Here ${}_s S_{l m \omega} (\theta)$ are the spin-weighted spheroidal harmonics \cite{Fackerell}, which obey the angular Teukolsky equation, while the radial functions ${}_s R_{lm \omega} (r)$ obey the radial Teukolsky equation. The angular and radial equations must be solved jointly for a given $(l,\,m)$ to yield the separation constants ${}_s A_{lm \omega}$ and frequencies $\omega_{lm}$. In what follows we suppress some or all of the indices $( l,\, m,\, \omega)$ where there is no danger of confusion. 

The angular Teukolsky equation is a Sturm-Liouville equation,
\begin{subequations}
\begin{align}
\label{eq:AngTeuk}
\csc \theta & \frac{d}{d\theta} \left( \sin \theta \frac{d {}_s S }{d \theta} \right) + V_\theta  {}_s S = 0 \,, \\
V_\theta  =&  a^2 \omega^2  \cos^2 \theta - m^2 \csc^2 \theta - 2 a \omega s \cos \theta \nonumber \\
&- 2 m s \cos \theta \csc^2 \theta - s^2 \cot^2 \theta + s + {}_s A_{lm} \,.
\end{align}
\end{subequations}
The rescaled radial function ${}_s u (r) = \Delta^{s/2} \sqrt{r^2+a^2} {}_s R(r)$ obeys the differential equation
\begin{subequations}
\begin{align}
\label{eqgeneralteuk}
& \frac{d^2 {}_su}{d r^2_*}+\left [\frac{K^2+2is(r-1)K+\Delta(4i\omega rs- {}_s\lambda_{lm\omega})}{(r^2+a^2)^2}\right ] {}_s u \notag \\
&- \left [G^2+\frac{d G}{d r_*}\right ] {}_s u=0,
\end{align}
which determines the complex eigenfrequencies $\omega$ after imposing boundary conditions. Here,
\begin{align}
\label{eqexplan}
G& =\frac{r \Delta}{(r^2+a^2)^2}+\frac{s(r-1)}{r^2+a^2}, \\
K&=-\omega (r^2+a^2)+am, \\ 
{}_s\lambda_{l m \omega} &={}_s A_{lm\omega}+a^2\omega^2-2am\omega,
\end{align}
\end{subequations}
and the tortoise coordinate $r_*$ is defined by
\begin{align}
\frac{dr_*}{dr} &\equiv \frac{r^2+a^2}{\Delta} \,.
\end{align}
Note that $r_* \to -\infty$ at the horizon $r_+$, and $r_* \to \infty$ as $r \to \infty$. 

We are interested in the study of QNMs, which are the eigenfrequencies of the homogeneous perturbation equations. To guarantee that the frequencies correspond to the physical oscillations of the Kerr black hole, the radial equation~\eqref{eqgeneralteuk} must be complemented with boundary conditions requiring only ingoing waves at the (future) event horizon and only outgoing waves at (future) null infinity. For the perturbations to decay in time, we see from Eq.~\eqref{eq:MasterSep} that the imaginary part of the frequency $\omega$ must be negative. We define the quantities $\omega_R$ and $\omega_I$ by
\begin{align}
\omega = \omega_R - i \omega_I , 
\end{align}
so that $\omega_I$ gives the decay rate of the QNM.

\subsection{Bound state formulation of the Teukolsky equation}
\label{sec:boundstate}
We discuss an alternative viewpoint for the solution of the radial Teukolsky equation, which improves our understanding of QNMs, especially in distinguishing DMs and ZDMs.  
As mentioned in the previous section, the radial Teukolsky equation describes a scattering problem, whose eigenvalues are the QNM frequencies when we enforce outgoing boundary conditions at infinity and ingoing boundary conditions at the horizon. It is useful to transform this scattering problem to a bound-state problem before applying the WKB approximation, as done by Mashhoon~\cite{mashhoon} for slowly rotating BHs. 
Here we describe a transformation procedure for perturbations of generic Kerr 
BHs (with arbitrary $a$) which is similar to Mashoon's for slowly rotating BHs. This procedure allows us to visualize the QNM wavefunctions, and gives us a new viewpoint from which to understand the behavior of the QNMs in terms of bound states in a potential well.
 
For generic Kerr BHs, the radial Teukolsky equation is given in 
Eq.~\eqref{eqgeneralteuk}. The angular eigenvalue can be expressed as a function of $L\equiv l+1/2$, $m$, and $\omega$: $A_{lm}=A(L,m,\omega,a)$. 
After the transformations
\begin{align}
\label{eq:transform}
 r& \rightarrow i r, & M &\rightarrow i M, & m &\rightarrow i m, & L &\rightarrow i L, \nonumber \\
 a &\rightarrow i a, & s &\rightarrow s, & \omega & \rightarrow \Omega,
 \end{align}
 the new radial equation becomes 
\begin{align}
\label{eqbound}
& \frac{d^2 u}{d r^2_*}-\left [\frac{\tilde{K}^2+2s(r-M)\tilde{K}-\Delta(4\Omega rs+\tilde{\lambda}^s_{lm})}{(r^2+a^2)^2}\right ] u \nonumber \\
&+\left [G^2+\frac{d G}{d r_*}\right ] u=0\,, \\
& \tilde{K}=\Omega(r^2+a^2)-ma,\\
& {}_s \tilde{\lambda}_{lm}=-A(iL,im,\Omega,ia)+a^2\Omega^2-2ma\Omega\,.
\end{align}
Note that we have restored the mass $M$ of the black hole, because it must also be transformed. 
The angular separation constant has the functional form
$A(L,m,\omega,a)=A(L,m/L,a\omega/L)$; then $A(iL,im,\Omega,ia)$ and 
${}_0 \tilde{\lambda}_{lm}$ are real if $\Omega$ is real. 
Now Eq.~(\ref{eqbound}) describes a bound state problem; when 
$r_* \rightarrow \pm\infty$, the wavefunction asymptotes to 
$e^{-\Omega r_*}$ or $e^{(\Omega-m\Omega_H)r_*}$. Here, $\Omega_H = a/2r_+$ is the horizon frequency, with $r_+ = 1 + (1 - a^2)^{1/2}$ denoting the radius of the outer horizon.
The eigenvalue $\Omega(L,m,n,a)$ should be a real-valued function 
depending on $L$, $m$, and $a$ ($n$ is the overtone number). 
Knowing the functional form of $\Omega$, we can then apply the inverse 
transform to obtain $\omega$:
\be
\omega=\omega_R-i\omega_I = \Omega(-iL,-im,n,-ia,-iM) \,.
\ee

We use this formulation in Sec.~\ref{sec:dmzdm} to plot ZDM wavefunctions in the extremal limit, and to discuss the DM frequencies in the eikonal limit.

\subsection{The eikonal limit and the WKB method\label{EikWKB}}
In the eikonal limit, where $l \gg 1$, we can find approximate solutions for the radial and angular Teukolsky equations. In the past, the eikonal method has been applied with great success to the Schwarzschild black hole \cite{Goebel1972,mashhoon} and to more generic spacetimes (see e.g. \cite{Cardoso:2008bp,Decanini:2010fz,Dolan10}). In this approximation, each eigenmode of the perturbation equations corresponds to an unstable photon orbit. 
The real part of the eigenmode frequency is given by the orbital frequency of the photon orbit, whereas the imaginary part is given by the Lyapunov exponent governing the divergence of null rays away from the unstable orbit.
 Similar studies applied the same techniques to the special cases of equatorial $(l = m)$ and polar $(m =0)$ perturbations of the Kerr spacetime~\cite{Dolan10}. 
In~\cite{Yang2012a}, approximate analytic solutions for the angular constants ${}_s A_{lm}$ and the QNM frequencies were found using WKB methods for any $m$, and these modes were again shown to be in correspondence with the unstable spherical orbits of the Kerr spacetime. 
We now review the WKB method and the approximate eikonal solutions.

We first define $L = l + 1/2$ ($\approx \sqrt{l(l+1)}$) and take the leading order terms in $L$ in the angular and radial Teukolsky equations, noting that ${}_s A_{lm} \sim O(L^2)$, $\omega_R \sim O(L)$, $\omega_I \sim O(1)$, and $m \lesssim L$. All terms involving the spin $s$ are subleading, and therefore we may ignore the spin-dependence of the variables and of the eigenvalues\footnote{There is a subtlety in that at the poles $(\theta = 0, \pi)$ there are terms involving the spin that diverge; however, it turns out that we can still neglect these terms, since when $m \neq 0$ the solution to Eq.~\eqref{eq:AngTeuk} vanishes at the poles, and for $m = 0$ we can rely on the known analytic solution discussed below.}. At leading order, the angular Teukolsky equation reduces to
\begin{subequations}
\label{eq:AngularWKBset}
\begin{align}
\label{eq:AngularWKB}
\frac{d^2 S}{dx^2} + \left(a^2 \omega^2 \sin^2 \theta \cos^2 \theta - m^2 + A_{lm} \sin^2 \theta\right) S \,,
\end{align}
where we have defined 
\begin{align}
x = \ln \left( \tan \frac \theta 2 \right ) \,, \qquad dx = \csc \theta d \theta .
\end{align}
\end{subequations}
The variable $x$ is analogous to the tortoise coordinate $r_*$ and is a key element, because it puts the angular Teukolsky equation in the standard form for a WKB treatment. Note that $x \to \pm \infty$ as $\theta \to 0, \pi$. We see that Eq.~\eqref{eq:AngularWKB} defines a bound state problem for $S$, where $S$ must decay to zero at infinity because the potential term asymptotes to $- m^2 $ as $x \to \pm \infty$ (for $m \neq 0$, see below). Using a bound-state WKB analysis, we find~\cite{Yang2012a} that the angular eigenvalues $A$ obey a Bohr-Sommerfeld quantization condition,
\begin{align}
\label{eq:BohrSom}
\int_{\theta_-}^{\theta_+} d \theta \sqrt{ a^2 \omega_R^2 \cos^2 \theta - m^2 \csc \theta^2 + A_{lm}}  = \pi (L - |m|),
\end{align}
where $\theta_\pm$ are the turning points in the potential, given by $\theta_+ = \pi - \theta_-$ and $\theta_- = \arcsin( \sqrt{3} -1)$. 
This is an integral condition for $A_{lm}$, given $\omega_R$. Such an integral condition can be jointly solved for $\omega_R$ and $A_{lm}$ numerically, once we apply another WKB approximation to find an expression for $\omega_R$. 
However, we can derive a useful result for $A$ by assuming $(a \omega_R/L)^2 \ll 1$ \cite{Yang2012a}. 
This assumption gives relatively accurate approximations in the Kerr spacetime, even in the case of nearly extremal black holes, since for modes with the horizon frequency we have $(a\omega_R/L)^2 \to 1/4$. 
Using this assumption, we can algebraically solve the integral condition~\eqref{eq:BohrSom}, keeping the first term in $(a \omega_R/L)^2$,
\begin{align}
\label{eq:AppxA}
A_{lm} \approx L^2 \left[ 1 - \frac{a^2 \omega_R^2}{2L^2} \left( 1 - \mu^2 \right) \right],
\end{align}
where $\mu \equiv m / L$.

\begin{figure*}[th]
\includegraphics[width=0.30\textwidth]{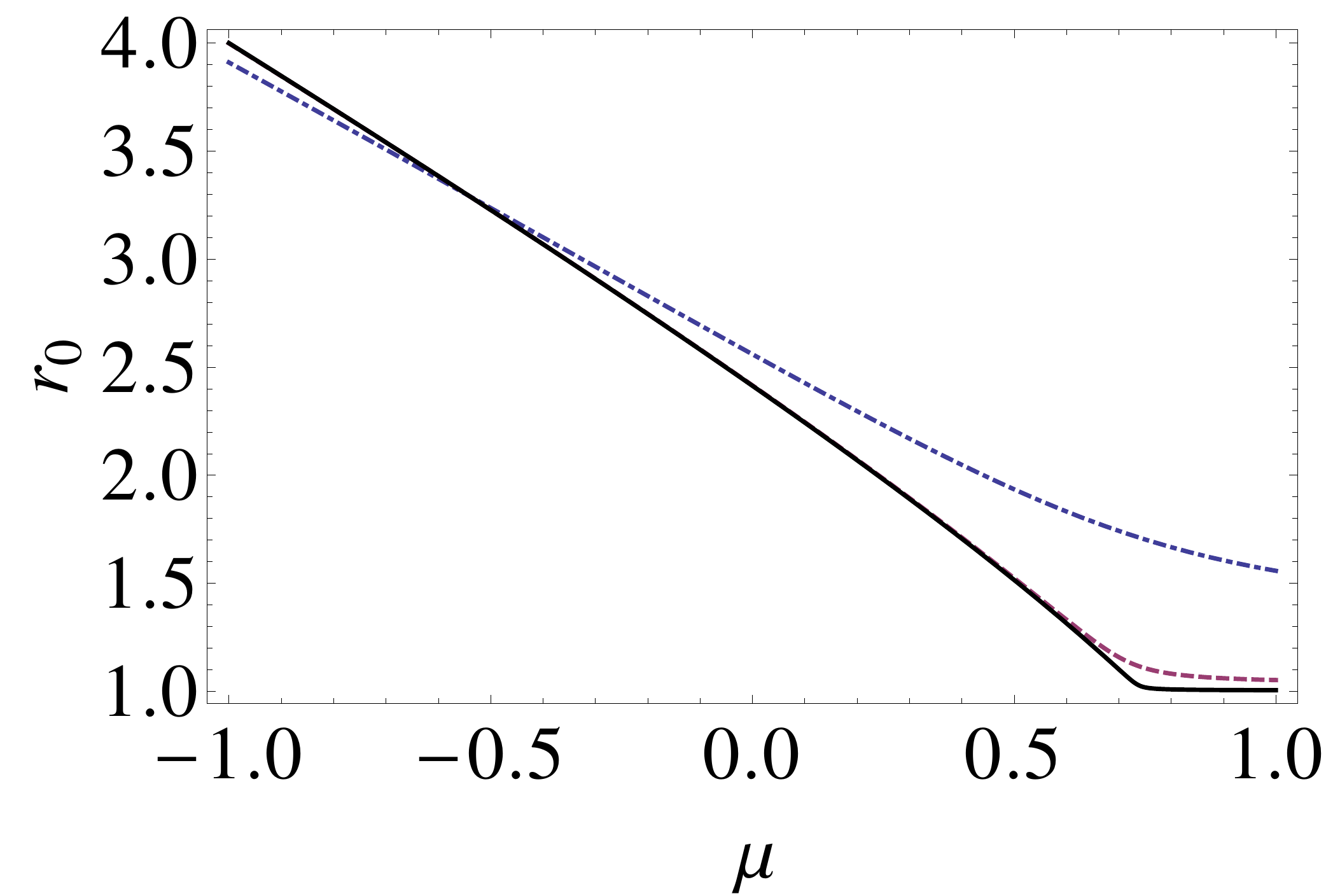}
\includegraphics[width=0.30\textwidth]{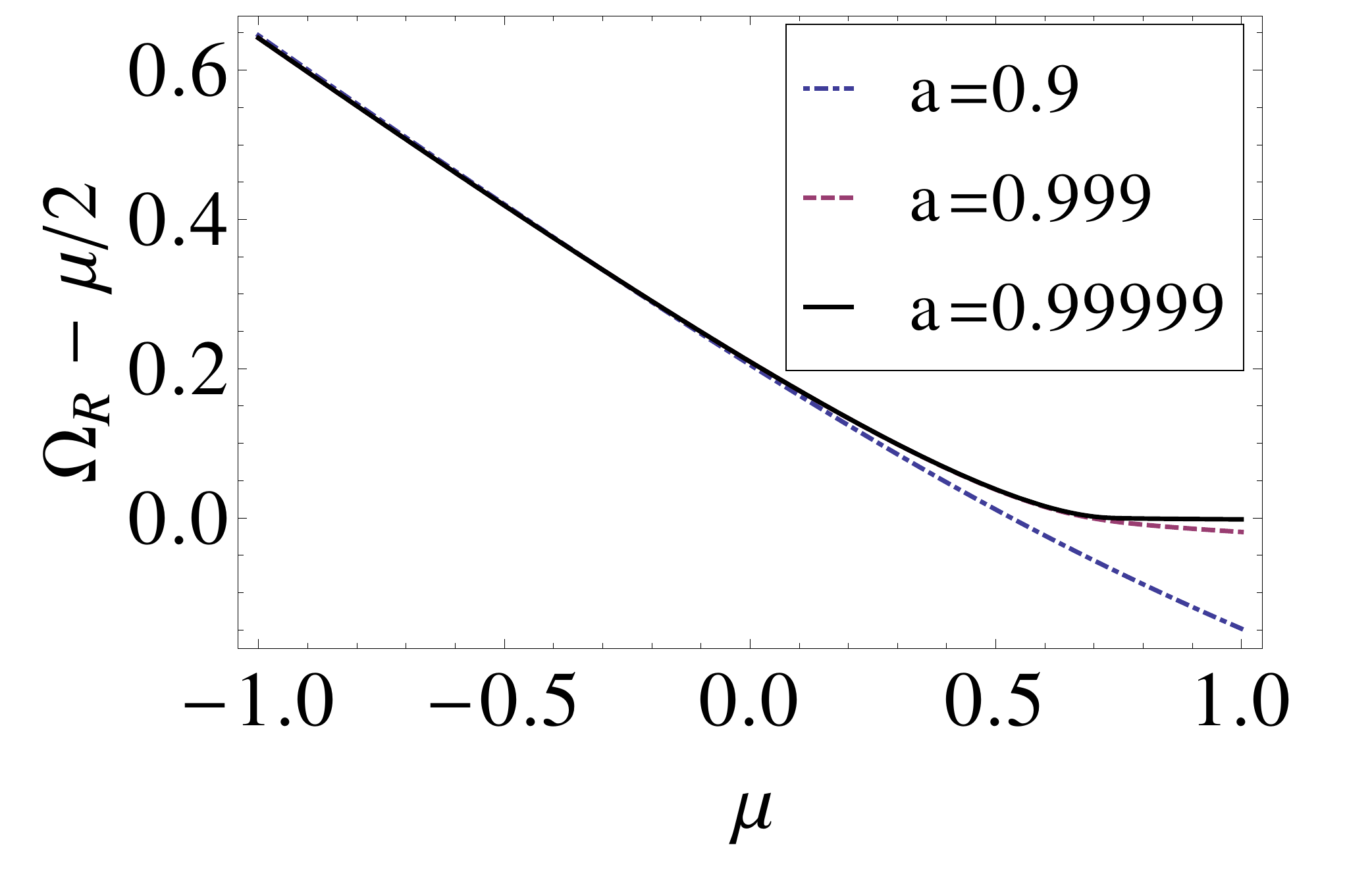}
\includegraphics[width=0.30\textwidth]{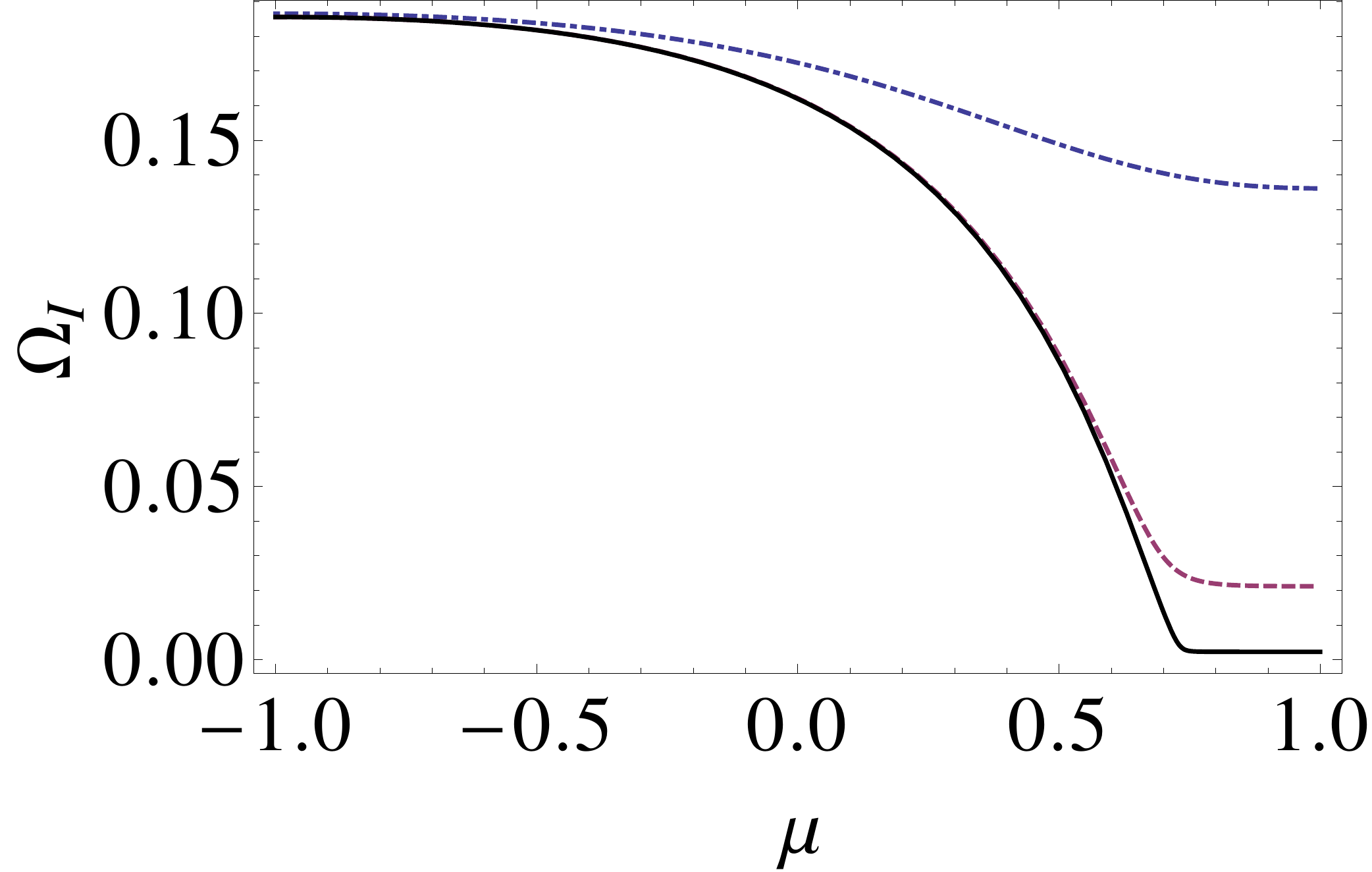}
\caption{WKB results for $r_0,\, \Omega_R - \mu/2$, and $\Omega_I$, for various values of $\mu$ and for increasing angular momenta. The angular momenta are $a = 0.9$ (blue, dashed-dotted lines), $a = 0.999$ (red, dotted lines), and $a = 0.99999$ (black, solid lines). Left: The radius $r_0$, which coincides with the horizon for $\mu_c < \mu <1$ as $a \to 1$. Middle: the scaled frequency minus the scaled angular frequency of the horizon, $\Omega_R - \mu/2$. Right: The decay rate $\Omega_I$, which vanishes for $\mu_c < \mu < 1$ as $a \to 1$.}
\label{fig:WKBplots}
\end{figure*}

Next, we apply the WKB approximation to the radial equation. In the eikonal limit, the radial equation becomes
\begin{align}
\label{eqr}
\frac{d^2 u}{d r^2_*}+V_r u = 0\,, \qquad V_r = \frac{K^2-\Delta \lambda}{(r^2+a^2)^2} \,.
\end{align}
Equation~\eqref{eqr} is more complicated than the standard Schr\"odinger-type equation, which has the form $ d^2 u /dx^2 -[V(x) - \omega^2] u = 0$. Nevertheless, we can draw on our intuition from familiar scattering problems. We adhere to common sign conventions in writing Eq.~\eqref{eqr}, but comparing it to Schr\"odinger-type equations leads us to refer to $-V_r$ as ``the potential.'' The potential generically has a ``peak'' $r_0$ where $-V_r$ attains its maximum, and we can perform a WKB expansion around it.
The real part of the frequency is given by the conditions that, at the peak, 
\begin{align}
V_r (r_0, \omega_R) = \left. \frac{\partial V_r}{\partial r} \right |_{r_0, \omega_R} = 0.
\end{align}
Using these conditions, we express $\omega_R$ in terms of the position of the peak,
\begin{equation}
\label{eq:OmegaR}
\omega_R = L \frac{(r_0-1) \mu a}{(3 - r_0)r_0^2 - (r_0 + 1)a^2}\,.
\end{equation}
The position of the peak $r_0$ is given by the roots of the sixth-order polynomial
\begin{align}
\label{eq:poly}
2r_0^4(r_0-3)^2 +4r_0^2[(1-\mu^2)r_0^2-2r_0-3(1-\mu^2)]a^2 \nonumber\\
+(1-\mu^2)[(2-\mu^2)r_0^2 + 2(2+\mu^2)r_0 + (2-\mu^2)]a^4\,.
\end{align}
The imaginary part of the frequency is given by the curvature of the potential at the peak through the equation 
\begin{align}
\label{eq:OmegaI}
\omega_I = (n + 1/2) \left. \frac{\sqrt{ 2 d^2 V_r/ dr_*^2}}{\partial V_r/\partial \omega}\right|_{r_0, \omega_R},
\end{align}
where $n$ gives the overtone number of the QNM. Defining $\Omega_R = \omega_R/L$, we have
\begin{widetext}
\begin{align}
\label{eq:EikonalDecay}
\omega_I &= (n + 1/2)\frac{ \Delta(r_0) \sqrt{4(6r_0^2 \Omega_R^2 - 1) + 2 a^2 \Omega^2_R(3 - \mu^2)}}{2 r_0^4 \Omega_R  - 4 a r_0 \mu + a^2 r_0 \Omega_R [r_0(3 - \mu^2) + 2 (1 + \mu^2)] + a^4 \Omega_R(1 - \mu^2)} \,.
\end{align}
\end{widetext} 
Equations~\eqref{eq:OmegaR},~\eqref{eq:poly}, and~\eqref{eq:EikonalDecay} rely on the assumption $(a \omega_R/L)^2 \ll 1$, but the WKB analysis can be carried out without this supplemental assumption~\cite{Yang2012a}, and the two methods agree to great precision. The QNMs with frequency $\omega$ correspond to unstable spherical photon orbits at radius $r_0$~\cite{Yang2012a}, with $m/L$ providing a measure for the inclination angle. For the case $m=0$, which corresponds to polar photon orbits, Eq.~\eqref{eq:OmegaR} becomes singular. In this case, $\omega_R$ can be written as
\begin{align}
\omega_R =\left. \pm L \frac {\pi \sqrt{\Delta}}{(r^2 + a^2 ) {\rm EllipE}\left[a^2 \Delta/(r^2 +a^2)^2 \right]} \right|_{r_p} \,,
\end{align}
where EllipE is the elliptic integral of the second kind and $r_p$ is the radius of the polar orbit, given by the roots of the cubic polynomial
\begin{align}
(r_p - 3) r_p^2 +(r_p + 1) a.
\end{align}
The real part of the frequency can be used in Eq.~\eqref{eq:EikonalDecay} and matches the eikonal approximation for the decay of the $m=0$ mode given in \cite{Dolan10}. Next, we discuss the behavior of these approximations as $a \to 1$.

\section{Nearly extremal Kerr}
\label{sec:NearlyExtreme}

In this section we examine the nearly extremal limit $\epsilon = 1 - a \ll 1$. First we gain insight by combining the nearly extremal approximation with the previously discussed eikonal limit. We then go beyond the eikonal limit by using matched asymptotic expansions.

\subsection{WKB results for nearly extremal Kerr}

As discussed in \cite{Yang2012a}, QNMs in the eikonal limit are clustered near the peak of the radial Teukolsky potential, which is essentially located at the radius of the spherical photon orbits (see also \cite{Cardoso:2008bp}). For NEK black holes, some of the spherical photon orbits reside near the horizon. The Lyapunov exponent, being proportional to the horizon surface gravity, is vanishingly small (zero if $a=1$). Correspondingly there is a family of QNMs that reside near the horizon, with imaginary parts $\omega_I$ which are very close to zero. 

In Figure~\ref{fig:WKBplots}, we plot the position of the peak $r_0$ and the real and imaginary parts of the rescaled WKB frequency $\Omega_R \equiv \omega_R/L$ and $\Omega_I = \omega_I/(n+1/2)$ over a range of $\mu\equiv m/L$ as $a \to 1$. Near the extremal limit, for some range of $\mu > 0$, $r_0$ approaches the horizon. The frequency $\Omega_R$ approaches $\mu$ times the horizon frequency for an extremal black hole ($\Omega_H \to 1/2$), therefore we plot the combination $\Omega_R - \mu/2$. At the same time the decay rate $\Omega_I$ falls to zero. 
The reason is that, for this set of modes, $d^2 V_r / d r_*^2$ scales as $\Delta^2$ in the extremal limit, and so as the peak of the potential approaches the horizon radius where $\Delta$ vanishes, the decay rate also vanishes. Further, the peak of the potential broadens, and the corresponding unstable photon orbits diverge from equilibrium more and more slowly.

\begin{figure}[t]
\includegraphics[width=1.0\columnwidth]{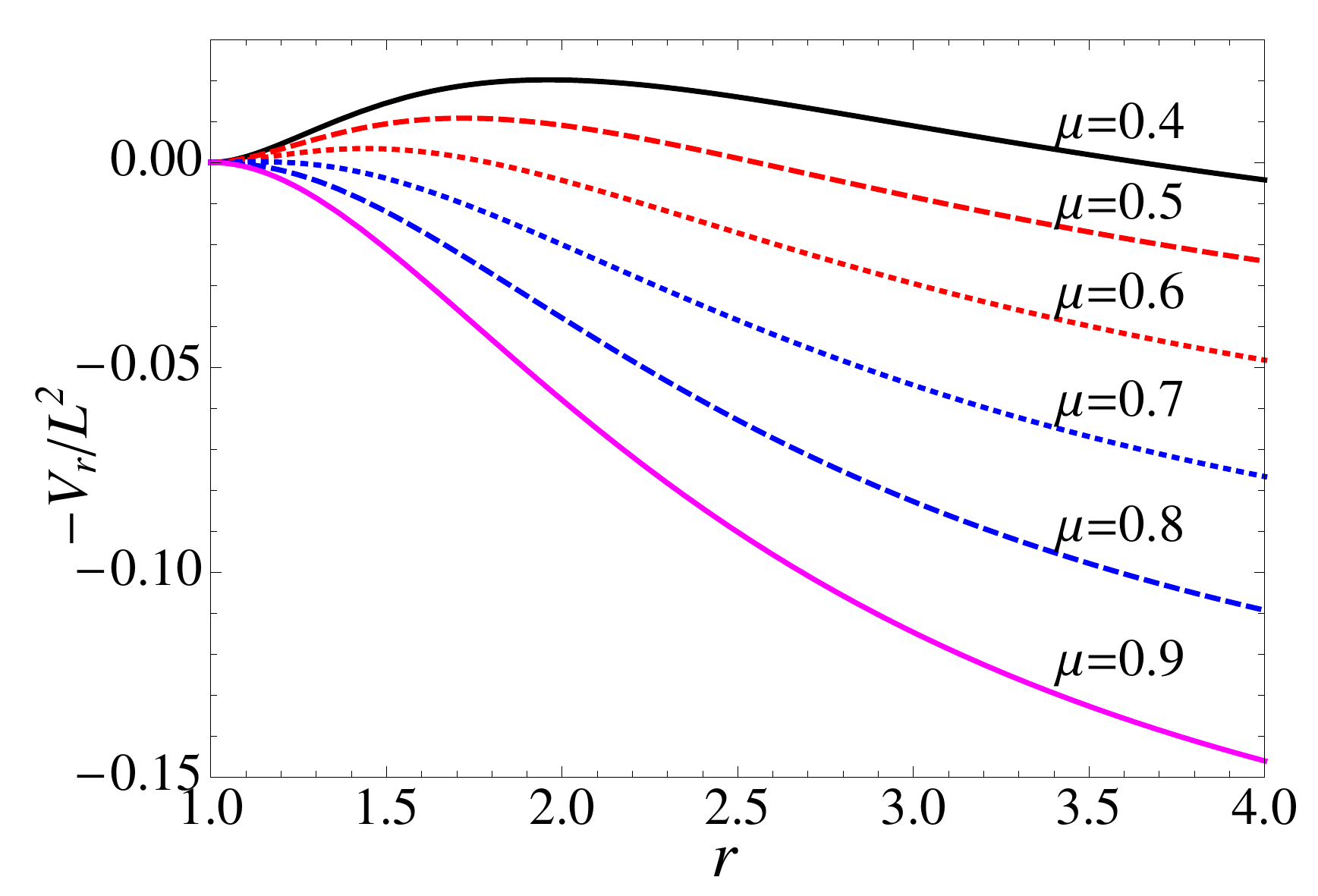}
\caption{Plot of the potential term Eq.~\eqref{eq:potential} for different $\mu$.
Here $\mu=0.4$, $0.5$, $ 0.6$, $0.7$, $0.8$, and $0.9$ correspond to black-solid, 
red-dashed, red-dotted, blue-dashed, blue-dotted, and magenta-solid curves, 
respectively.
The transition happens between $0.7$ to $0.8$. Reproduced from~\cite{Yang2012b}.}
\label{fig:pplot}
\end{figure}

However, there is still the question of why the decay vanishes only above some critical value of $\mu$, which we call $\mu_c$. In the extremal limit, the potential is
\begin{equation}
\label{eq:potential}
-V_r=-L^2\frac{(r-1)^2}{(r^2+1)^2}\left [ \frac{(r+1)^2}{4}\mu^2-\alpha+\frac{3}{4}\mu^2\right ] ,
\end{equation}
where 
\begin{align}
\alpha = \frac{A_{lm}}{L^2} = 1 - \frac{a^2 \Omega_R^2}{2} \left(1 - \mu^2 \right).
\end{align}
 In Fig.~\ref{fig:pplot}, which we reproduce from~\cite{Yang2012b}, we plot $- V_r/L^2$ from Eq.~\eqref{eq:potential} for a few values of $\mu$. For $\mu < \mu_c$, the peak of the potential stays outside the horizon and there is a second extremum of the potential at the horizon. As $\mu$ increases, this peak broadens and approaches the horizon. Beyond $\mu_c$, there is only one peak at the horizon. The WKB picture associates the approximate eigenfrequencies and wavefunctions with the peak. When the peak remains outside the horizon, there are frequency modes associated with it that retain finite decay. The existence of an extremum at the horizon for all $\mu$ already hints that there are modes which reside at the horizon that the WKB expansion about the peak misses, as we show in Sec.~\ref{sec:soldet}. 

In~\cite{Yang2012a}, the approximate value $\mu_c \approx 0.74$ was given by inspection of the behavior of $\Omega_I$ and $r_0$. Figure~\ref{fig:pplot} demonstrates that $\mu_c$ may be found by ensuring that there are no peaks of the potential outside the horizon, which gives the requirement that for $\mu \geq \mu_c$,
\begin{equation}
\label{eqalpha}
 \frac{(r+1)^2}{4}\mu^2-\alpha+\frac{3}{4}\mu^2 >0\quad {\rm for} \quad r=1\,.
 \end{equation}
Alternatively, if we define 
\begin{align}
\label{eq:DefineF}
\mathcal F_0 = L \sqrt{\frac{7 \mu^2}{4} - \left. \alpha \right|_{\Omega_R = \mu/2}}\,,
\end{align}
then the requirement reads $\mathcal F_0^2 > 0$. Hod~\cite{hod3} used the approximation~\eqref{eq:AppxA} for $A_{lm}$ and found that $\mu_c \approx [ ( 15 - \sqrt{193})/2 ] ^{1/2} \approx 0.744$. The ``exact'' (in the eikonal limit) value for $\mu_c$ can be found using the Bohr-Sommerfeld condition~\eqref{eq:BohrSom}, once the condition $\alpha( \mu_c) = 7 \mu^2/4$ is used to eliminate $A_{lm}$~\cite{Yang2012b}.

With some insight into the behavior of the eikonal QNMs for nearly extremal angular momenta, we can also compute simple expressions for the nearly extremal WKB modes in the case where the peak approaches the horizon. We define $\epsilon = 1- a \ll 1$ as a small parameter, and note that $r_+ = 1 + \sqrt{2 \epsilon}$ to leading order in the nearly extremal limit. If we assume that the peak approaches the horizon at a similar rate, $r_0 = 1 + c \sqrt{\epsilon}$, and solve the WKB equations for $r_0$ and $\omega$, we find to leading order
\begin{align}
r_0 =& 1 - \frac{m \sqrt{2 \epsilon}}{\mathcal F_0} , \\
\omega =& \left( \frac m 2 - \mathcal F_0 \sqrt{\frac{\epsilon}{2}} \right)- i \left(n + \frac 12\right) \sqrt{\frac{\epsilon}{2}} .
\end{align}
Our WKB results for perturbed NEK black holes indicate two distinct phases of QNMs in the eikonal limit $l \gg 1$: for $\mu < \mu_c$, the peak of the potential remains outside the horizon and the WKB frequency modes have finite decay; for $\mu > \mu_c$, the peak is at the horizon and the modes have vanishing decay. 
To gain a better understanding of the NEK spacetime, we must go beyond the eikonal limit. An expansion in $\epsilon$ provides the approximate solution to the radial Teukolsky equation for any $(l, \, m)$. We now review this solution and the corresponding QNM frequencies.

\subsection{Matched asymptotic expansions}
\label{sec:matchex}

To derive a formula for the QNM frequencies for the NEK black hole ($\epsilon \ll1$), we solve the radial Teukolsky equation in the two asymptotic domains far from and near to the horizon, and then we match these solutions together in an intermediate region, as demonstrated first by Teukolsky and Press~\cite{TeukolskyPress3}.
The long-lived nature of the QNMs found by this method was first noted by Detweiler~\cite{detweiler}. His results were expanded upon by other authors~\cite{sasaki,anderson,cardoso}, although using an incorrect assumption about the scaling of near-extremal quantities. The correct QNM expressions we derive here were first found by Hod~\cite{hod2}, and have been rediscovered in the context of the Kerr/CFT duality~\cite{Dias} within the near-horizon extremal Kerr spacetime~\cite{bardeen}. 
In addition to reviewing the matched asymptotic expansion method in a manner slightly different than the preceding literature, we discuss (to our knowledge, for the first time) the range of validity of our approximations and the leading errors on the frequencies. 

First we construct the solution far away from the horizon by expanding the radial Teukolsky equation in large $r$ and small $\epsilon$, with $(r-1) \gg \sqrt{\epsilon}$. We substitute $u = \Delta^{s/2}\sqrt{r^2 +a^2} \, R$ into the radial Teukolsky equation~\eqref{eqgeneralteuk}, and write it as 
\begin{align}
\label{eq:ConfDE}
&x^2  R'' + 2(s+1) x R' \notag \\
&+ \left[ \omega^2 (x + 2)^2 +2 i \omega s x - \lambda \right] R= 0 ,
\end{align}
where $x = r -1$. The solutions to this equation can be found in terms of confluent hypergeometric functions:
\begin{align}
\label{eqconfsln}
R = & A\, e^{-i \omega x} x^{-1/2 - s + i \delta} \notag \\
& \times {}_1 F_1(1/2-s + i\delta + 2 i \omega, 1 + 2 i \delta, 2 i \omega x) \notag \\
&+ B\, (\delta \to - \delta)\,,
\end{align}
where $\delta^2 \equiv 7m^2/4-(s+1/2)^2-A_{lm}$, and where $(\delta \to - \delta)$ indicates that we should replace $\delta$ with $-\delta$ in the preceding functions. We use the convention that $\Re(\delta) \geq 0$ when $\delta^2 $ is positive and $\Im (\delta) \geq 0$ when $\delta^2$ is negative. Taking the limit $x \to \infty$ and requiring no incoming waves (to find QNM solutions), we find that the coefficients $A$ and $B$ must obey\footnote{Here the appropriate solution to Eq.~\eqref{eq:ConfDE} has the outer boundary condition ${}_1 F_1 (a,b,z) \to \Gamma(b) [ e^z z^{a-b}/\Gamma(a) + (-z)^{-a}/\Gamma(b-a)]$ as $z = 2 i \omega \to \infty$; the former term contributes to outgoing solutions for $R$, while the latter provides the ingoing waves and must be canceled by the conditions on $A$ and $B$. As $z \to 0$ this solution has ${}_1 F_1(a,b,z) \to 1$.}
\begin{align}
\label{eqHorizonBC}
\frac{A}{B} =&  e^{\pi \delta + 2 i \delta \ln(2 \omega)} \frac{\Gamma(-2 i \delta) \Gamma( 1/2 - s + i \delta - 2 i \omega)}{\Gamma(2 i \delta) \Gamma( 1/2 -s - i \delta - 2 i \omega)}.
\end{align}

Next, we solve the radial Teukolsky equation in the near-horizon limit, with $(r -r_+)/r_+ \ll 1$. In this case we also require that the frequencies are near the critical frequency for superradiance, $m \Omega_H$, so that $\omega - m\Omega_H \ll 1$. To this end, we define the rescaled frequency difference
\begin{align}
\label{eq:tildeomega}
\tilde \omega = \frac{\omega - m \Omega_H}{\sqrt{\epsilon}},
\end{align}
and seek solutions where $\tilde \omega$ is order unity. Substituting $\tilde \omega$ into the radial Teukolsky equation in the near-extremal and near-horizon limits allows us to write Eq.~\eqref{eqgeneralteuk} in a self-similar form 
\begin{align}
\label{eq:nearhorizon}
y^2 u'' &+ y u' + V_y u = 0, \\
V_y =& \left( \tilde \omega - \frac{i s}{\sqrt{2}} \right)^2 + \frac{2(\sqrt{2}\tilde \omega - m)(m-is)y}{1-y} 
\notag \\ & 
+\frac{2[\mathcal F_0^2 - s(s+1)]}{(1-y)^2}
\end{align}
where we set $y = e^{\sqrt{2 \epsilon} r_*}$. The solution to this equation can be written in terms of the hypergeometric function, 
\begin{align}
\label{eq:hyper1}
u = y^{-p} (1-y)^{-q} {}_2 F_1 (\alpha,\beta,\gamma, y),
\end{align}
with
\begin{align}
&p = i \bar \omega/\sqrt{2}, &\qquad  q  = -1/2 - i \delta,  \notag \\
&\alpha = 1/2 + i (\bar m + \delta-  \sqrt{2} \bar \omega), & \qquad \beta = 1/2 - i \bar m + i \delta, \notag \\
& \gamma =  1 - i \sqrt{2} \bar \omega,
\end{align}
where we have chosen the signs of $p$ and $q$ so that the solution satisfies the desired boundary conditions. We have also defined $\bar \omega = \tilde \omega - i s / \sqrt{2}$ and $\bar m = m -i s$. By recalling that $y\to 0$ and ${}_2 F_1 (\alpha,\beta,\gamma,y) \to 1$ as $r \to r_+$, we see that this solution behaves like an ingoing wave at the horizon, $u \propto e^{- i (\omega - m \Omega_H) r_*}$. 

Finally, we match our two asymptotic solutions in an intermediate region where $\sqrt{\epsilon} \ll x \ll 1$. We accomplish this by defining $z = 1 - y$ and taking the limit of the inner solution as $z \to 0$ ($y \to 1$, where there is a pole in the hypergeometric function). The inversion of ${}_2F_1$ gives~\cite{nist}
\begin{widetext}
\begin{align}
{}_2F_1(\alpha,\beta,\gamma,y) &= \frac{\Gamma(\gamma)\Gamma(\gamma - \alpha -\beta)}{\Gamma(\gamma - \alpha) \Gamma (\gamma - \beta)}{}_2F_1(\alpha,\beta,\alpha + \beta - \gamma +1,z) \notag \\
&+ \frac{\Gamma(\gamma)\Gamma(\alpha + \beta - \gamma)}{\Gamma(\alpha) \Gamma(\beta)} z^{\gamma - \alpha - \beta} {}_2F_1(\gamma- \alpha,\gamma -\beta,\gamma - \alpha - \beta+1, z),
\end{align}
and this is set equal to the $x \to 0$ limit of the solution~\eqref{eqconfsln}, using $u \approx x^s R$ and $z \approx \sqrt{8 \epsilon}/x$. In the matching region, $R = A x^{-1/2 - s + i \delta} + B (\delta \to - \delta)$. We equate the exponents of $x$, which allows us to solve for $A$ and $B$. The expressions are given in Appendix~\ref{ap:NEKCalcs}. From this, we have 
\begin{align}
\label{eqmatch}
\frac{A}{B} & = e^{- i \delta\ln(8 \epsilon)} \frac{\Gamma(2 i \delta)\Gamma(1/2 + i  m -i \delta -i \sqrt{2} \tilde \omega) \Gamma( 1/2 + s - i m - i \delta )}{\Gamma(-2 i \delta)\Gamma(1/2 + i m + i \delta - i \sqrt{2} \tilde \omega) \Gamma(1/2 + s - im + i\delta) }.
\end{align}
Together, Eqs.~\eqref{eqHorizonBC} and~\eqref{eqmatch} yield a condition on the NEK frequency modes,
\begin{align}
\label{eqdetweiler}
e^{-\pi \delta-2 i \delta \ln(m)-i\delta \ln(8\epsilon)}\frac{\Gamma^2(2 i \delta)\Gamma(1/2+s-i m -i \delta)\Gamma(1/2-s-i m -i \delta)\Gamma[1/2+i(m-\delta-\sqrt{2}\tilde{\omega})]}{\Gamma^2(-2 i \delta)\Gamma(1/2+s -i m +i \delta)\Gamma(1/2-s-i m +i \delta)\Gamma[1/2+i(m+\delta-\sqrt{2}\tilde{\omega})]}
=1 .
\end{align}
\end{widetext}
This resonance condition determines the allowed values of $\tilde \omega$, and thereby the QNM frequencies $\omega$. We solve it by expanding the Gamma function in the numerator around its poles, as described next.

\subsection{Solutions of the resonance condition}
\label{sec:soldet}

With our conventions on $\delta$, the left-hand side of Eq.~\eqref{eqdetweiler} is generally a very small number. The only way to satisfy the equality is to be near one of the poles of the Gamma functions in the numerator, and this condition determines the value of $\tilde \omega$. When $m \geq 0$, we can find such a solution near the poles at the negative integers,
\begin{align}
\label{eq:PoleSln}
\tilde \omega = \frac{m - \delta}{\sqrt{2}} + \eta - \frac{i}{\sqrt{2}} \left(n + \frac 12\right).
\end{align}
The factor $\eta$ is the shift of the scaled frequency $\tilde{\omega}$ from its value at the Gamma function's pole. The shift $\eta$ is generally very small, and can be taken to be zero except in a few specific cases. We analyze its properties by inserting the frequency~\eqref{eq:PoleSln} into the condition~\eqref{eqdetweiler}, and expanding the relevant Gamma function near its poles.

\begin{figure}[b]
\includegraphics[width=0.53\textwidth]{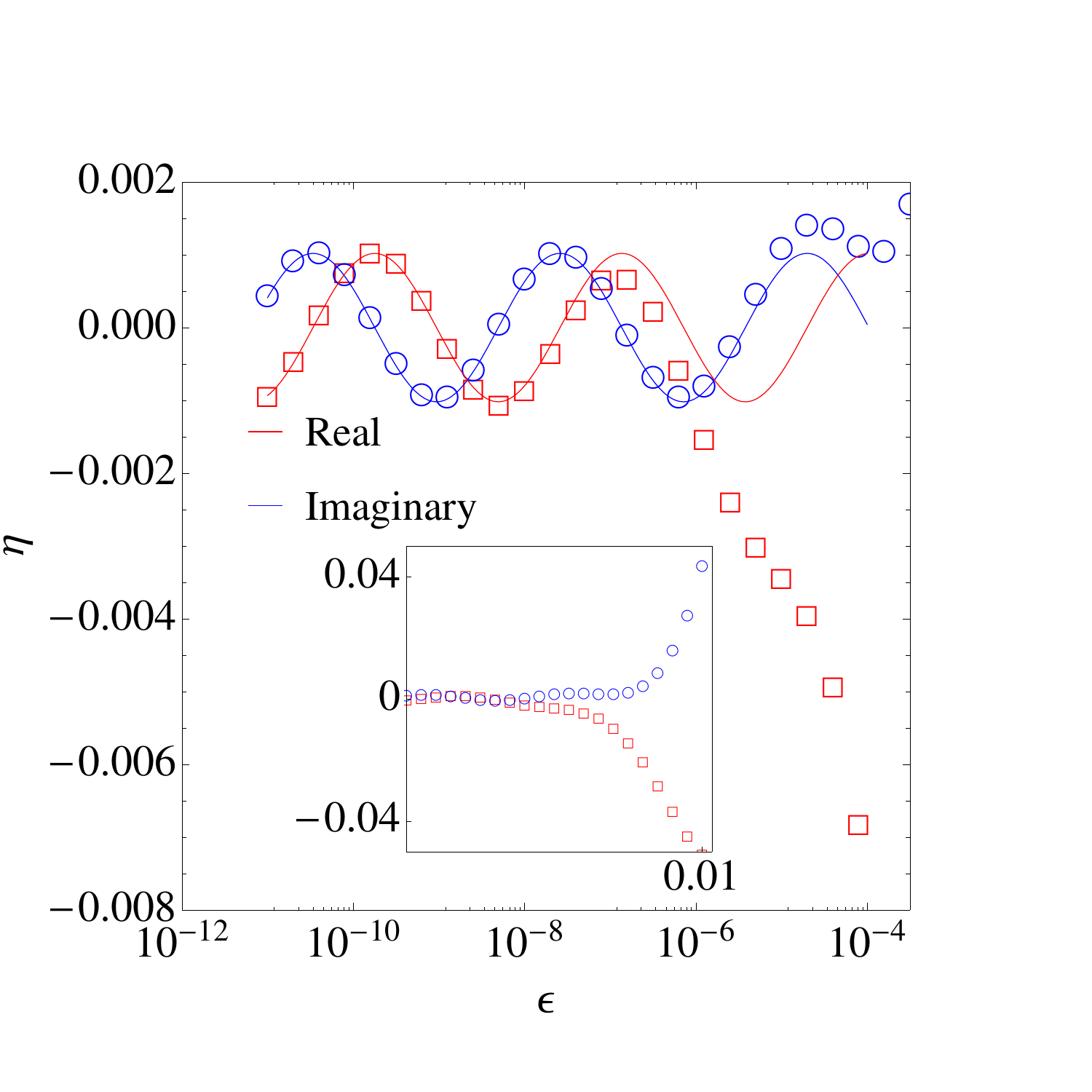}
\vspace{-6mm}
\caption{Comparison between numerical (circles and squares) and analytical (solid lines) calculations for the $a\rightarrow 1$ behavior of the $(l,m)=(2,2)$ mode. Here red square (blue circle) stands for the real (imaginary) part of $\eta$. The inset shows that, up to $\epsilon=0.01$, the real and imaginary parts of $\eta$ are still bounded by $\pm0.04$. 
\label{fig:qnm22}}
\end{figure}

For example, when $\delta^2>0$ and $|\delta| \sim 1$, Eq.~(\ref{eqdetweiler}) implies
\begin{align}
\label{eqeta}
\eta=&e^{-i\phi}\frac{e^{-\pi\delta}}{\sqrt{2}n!} \left|\frac{\Gamma(1/2+s-i m -i \delta)}{ \Gamma(-n+2i\delta) \Gamma(1/2+s-i m +i \delta)} \right| \notag \\ 
&\times \left |\frac{\Gamma(1/2-s-i m -i \delta)}{\Gamma(1/2-s-i m +i \delta)} \right|,
\end{align}
where the phase $\phi$ contains the factor $\delta \ln(8\epsilon)$, and we have
used the approximation
\begin{align}
\Gamma( - n -i \sqrt{2}  \eta) \approx (-1)^n/[n! (-i\sqrt{2}\eta)]\,,
\end{align} 
which is derived from the relations $\Gamma (1+ z) = z \Gamma (z)$ and 
$\Gamma(-i\sqrt{2}\eta) \approx 1/(-i \sqrt{2}\eta)$ for small $\eta$. 
It is clear from Eq.~\eqref{eqeta} that $\eta$ oscillates sinusoidally in 
$\ln{\epsilon}$. 
As we illustrate in Fig.~\ref{fig:qnm22} for the $(l,m)=(2,2)$ mode, 
this oscillation is in good agreement with the numerical results for the QNM frequencies using Leaver's method~\cite{leaver}. Figure~\ref{fig:qnm22} also demonstrates that for this mode, $\eta$ remains small even for $\epsilon = 10^{-2}$.

\begin{figure}[t]
\includegraphics[width=0.43\textwidth]{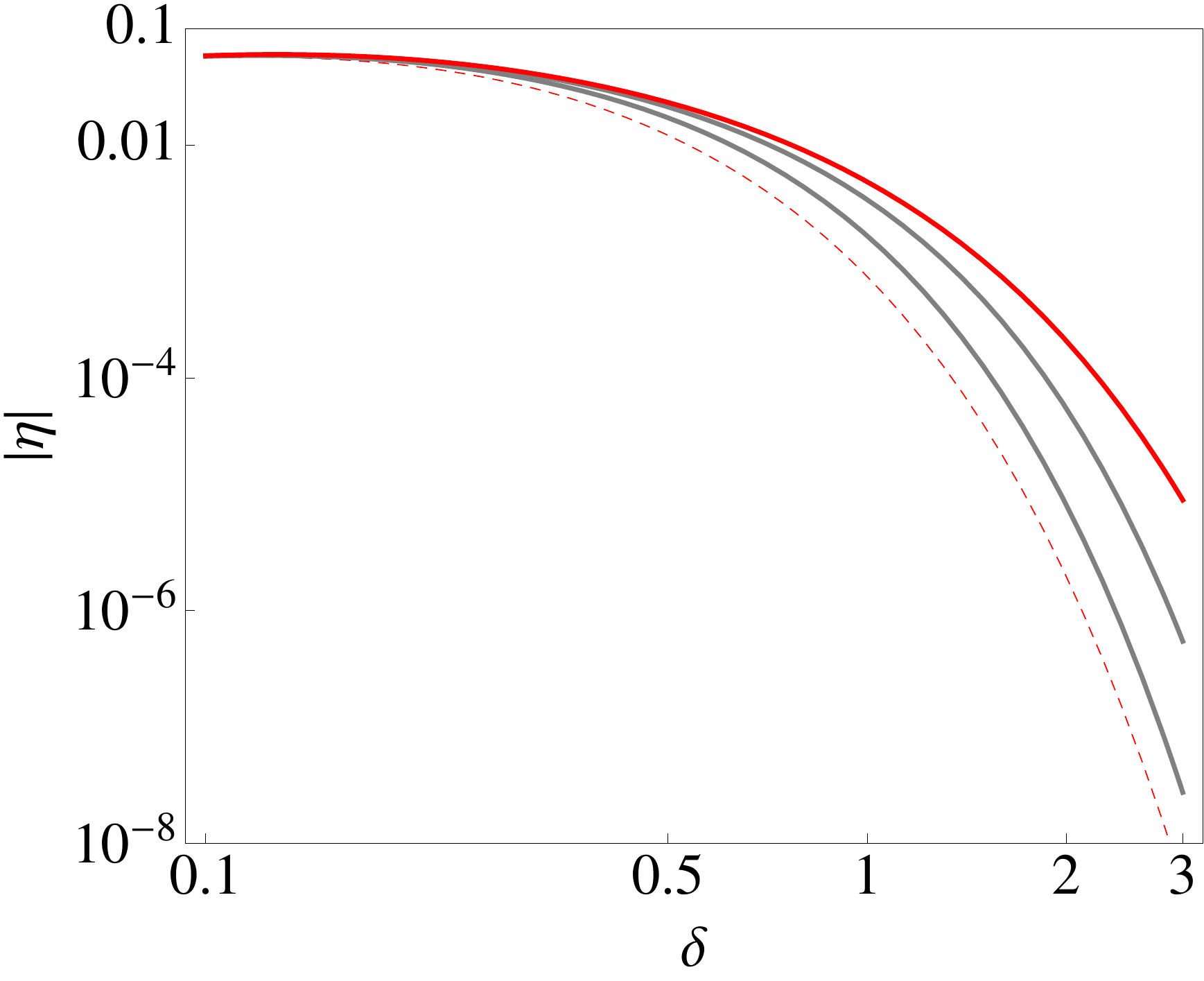}
\caption{Magnitude of $\eta$ as a function of $\delta$. The red dashed line is for $n=0$, given by the lower bound in Eq.~(\ref{eqetan0}). The two gray lines are for 
$n=1$ and $n=5$. The solid red line corresponds to $n\rightarrow \infty$, and is given by the upper bound of Eq.~(\ref{eqetabound}).
\label{fig:abseta}}
\end{figure}

The behavior of $\eta$ depends strongly on $\delta$. 
For modes with $\delta^2>0$ that are near the boundary between the single- and 
double-phase regimes, $\delta$ can be order of unity, and $m$ is generally 
larger than $\delta$. 
This means it is possible to make the approximation
\be
\frac{\Gamma(1/2\pm s-im-i\delta)}{\Gamma(1/2\pm s-im+i\delta)}\approx e^{-\pi\delta},
\ee
and we find
\be
\eta \approx e^{-i\phi}\frac{e^{-3\pi\delta}}{\sqrt{2}\Gamma(n+1)|\Gamma(-n+2i\delta)|}.
\ee
Using the fact that
\be
|\Gamma(2i\delta)|=|\Gamma(-2i\delta)| 
=\sqrt{\frac{\pi}{2\delta {\rm sinh(2\pi\delta)}}}\,,
\ee
for $n=0$, we have
\be\label{eqetan0}
|\eta|=\frac{e^{-3\pi\delta}}{\sqrt{\pi}}\sqrt{\delta {\rm sinh}\,2\pi\delta}.
\ee
For larger $n$, we can put the following bounds on $\eta$:
\be\label{eqetabound}
e^{-3\pi\delta}\sqrt{\frac{\delta {\rm sinh}\,2\pi\delta}{\pi}}<|\eta|<e^{-3\pi\delta}\frac{{\rm sinh}\,2\pi\delta}{\sqrt{2}\pi}\,.
\ee
We see that $\eta$ remains small whatever the value of the overtone number $n$ is, so that the ZDM formula holds for large $n$, even for $0<\delta<1$. So long as $\epsilon$ is small enough that $|\omega - m/2| \ll 1$, there are an arbitrarily large number of ZDM overtones, and all their frequencies remain separated from the horizon frequency by terms of order $\sqrt{\epsilon}$. We plot $\eta$ as a function of $\delta$ for various values of $n$ in Fig.~\ref{fig:abseta}. For $\delta >1$, the right-hand side of Eq.~(\ref{eqeta}) is very small, and these modes are very close to the horizon frequency, with decay rates proportional to $n+1/2$.

The other case we need to consider is $\delta^2 <0$. 
When $|\delta| \gtrsim 1$, the right-hand side of Eq.~\eqref{eqeta} decays 
rapidly as $|\delta|$ grows. 
However, when $\delta^2 < 0$ and $|\delta|$ is small, $\eta $ can become large.
For imaginary $\delta$, the factor $\delta \ln (8 \epsilon)$ in the phase 
of~\eqref{eqeta} suppresses the amplitude, giving 
$\eta \propto \epsilon^{|\delta|}$; thus, even though $\eta$ can be large while
$\epsilon \ll 1$, $\eta$ still vanishes in the limit $\epsilon \to 0$, and it is small for sufficiently small $\epsilon$.

Another set of solutions for the condition~\eqref{eqdetweiler} seems possible, near the poles of $\Gamma(2i\delta)$. In fact, we can show that this case invalidates the matching procedure used to derive Eq.~\eqref{eqdetweiler} in the first place. Suppose $2i\delta\approx -n$, where $n$ is a positive integer. 
Then there are two nearly degenerate hypergeometric function solutions of the 
near-horizon Teukolsky equation:
\begin{equation}
R(r)={}_2F_1\left (\alpha_{\pm},\beta_{\pm},1+s-i\sqrt{2}\tilde{\omega},-\frac{r}{2\sqrt{2\epsilon}}\right )\,,
\end{equation}
\begin{align}
\alpha_+&=-im+s+\frac{1}{2}+i\delta, &\alpha_-=-im+s+\frac{1}{2}+i\delta+n, \nonumber \\
\beta_+&=-im+s+\frac{1}{2}-i\delta, &\beta_-=-im+s+\frac{1}{2}-i\delta-n .
\nonumber
\end{align}
These two solutions are nearly degenerate because $\alpha_+\approx\beta_-$ and 
$\alpha_-\approx\beta_+$. To construct the Teukolsky solution in the near-horizon regime, both solutions 
have to be taken into account. In this case, the asymptotic matching procedure presented previously does not apply, and the condition for the QNM frequencies is no longer given by Eq.~\eqref{eqdetweiler}. We looked for these poles numerically, using Leaver's method. Our numerical investigations indicate that there are no QNMs where $2 i\delta \approx -n$. The numerical methods used here and in later sections are discussed in more detail in Appendix~\ref{ap:NumericalMethods}.

\subsection{DMs and ZDMs}
\label{sec:dmzdm}

We have seen that, except in a small range of $\delta^2 <0$ and $|\delta| \sim 1$, the shift in the scaled frequency $\eta$ is small, and we expect QNM solutions with frequencies
\be
\label{eqhod2}
\omega \approx \frac{m}{2}-\frac{\delta \sqrt{\epsilon}}{\sqrt{2}}-i\left (n+\frac{1}{2}\right ) \frac{\sqrt{\epsilon}}{\sqrt{2}} .
\ee
Except for terms that depend on the spin $s$ of the mode, $\mathcal F_0$ and $\delta$ are the same. In fact, if we define a more general quantity $\mathcal F_s$ via
\begin{align}
\mathcal F_s^2 = \frac{7 m^2}{4} -s(s+1) - {}_s A_{lm} ,
\end{align}
then $\mathcal F_s$ and $\delta$ are identical, up to a factor of $1/4$. As such, the question whether $\mu$ is above or below its critical value $\mu_c$ is essentially the same as the question whether $\delta^2$ is positive or negative. The frequency solutions~\eqref{eqhod2} exist also for negative $\delta^2$. These frequencies match our WKB results for $\mu > \mu_c$, but for $0 < \mu < \mu_c$ they still predict ZDMs, indexed by an overtone number $n$, while the WKB results give DMs.
There are no counter-rotating ZDMs $(m <0)$ because the matched asymptotic expansion requires a small frequency difference $\omega - m \Omega_H$, which does not occur for a positive $\Re[\omega]$ and negative $m$. Meanwhile, the symmetry $\omega_{lm} = - \omega^*_{lm}$ gives ZDMs with both $\Re[\omega]<0$ and $m<0$. Note also that the overtone number $n$ of the ZDMs is not precisely the same as the overtone number of Kerr QNMs at arbitrary angular momenta, as we see later.

It turns out that the {\it bound-state} formulation of the Teukolsky equation discussed in Sec.~\ref{sec:boundstate} gives us an intuitive way to understand DMs and ZDMs. We first discuss them in the eikonal limit, and we start with the DMs. In this case, $A(iL,im,\Omega,ia)=-A(L,m,\Omega,a)$ 
and Eq.~(\ref{eqbound}) 
becomes
\be
\frac{d^2 u}{d r^2_*}-\left [\frac{\tilde{K}^2-\Delta (A+a^2\Omega^2-2am\Omega)}{(r^2+a^2)^2}\right ] u=\frac{d^2 u}{d r^2_*}-V_r u=0\,.
\ee
We keep only leading-order terms in $L$. 
Because the potential well is very deep as $L \gg 1$, the fundamental mode and 
the first few overtones should be located near the bottom of the potential 
well: $V_r(\Omega_0,r_{\rm peak})=\partial_r V_r(\Omega_0,r_{\rm peak})=0$. 
We Taylor expand $V_r$ near its extrema
\be
V_r(\Omega_0+\delta \Omega,r_{* \rm peak}+\delta r_*)\approx \partial_{\Omega} V_r \delta \Omega+\frac{1}{2}V''_r \delta r^2_*\,,
\ee
where primes denote derivatives with respect to $r_*$. 
The new equation becomes
\be
\frac{d^2 u}{d r^2_*}=\left (\partial_{\Omega} V_r \delta \Omega+\frac{1}{2}V''_r \delta r^2_* \right )u\,.
\ee
This is now a standard bound-state eigenvalue problem, with solution
\be
\delta \Omega =-\left ( n+\frac{1}{2}\right ) \frac{\sqrt{2V''_r}}{\partial_{\Omega}V_r}\,,
\ee
or
\begin{align}
\Omega&=\Omega_0-\left ( n+\frac{1}{2}\right ) \frac{\sqrt{2V''_r}}{\partial_{\Omega}V_r} \nonumber \\
&=\frac{1}{M}\left [L\omega_0(\mu)-\left (n+\frac{1}{2}\right )\omega_1(\mu)\right]\,.
\end{align}
We obtain
\begin{align}
\omega &=\Omega(-iL,-im,n,-ia,-iM)\nonumber \\
&=\Omega_0-i \left ( n+\frac{1}{2}\right ) \frac{\sqrt{2V''_r}}{\partial_{\Omega}V_r}\,,
\end{align}
which agrees with the eikonal limit QNM formula in \cite{Iyer}.

\begin{figure}[t!]
\includegraphics[width=0.9\columnwidth]{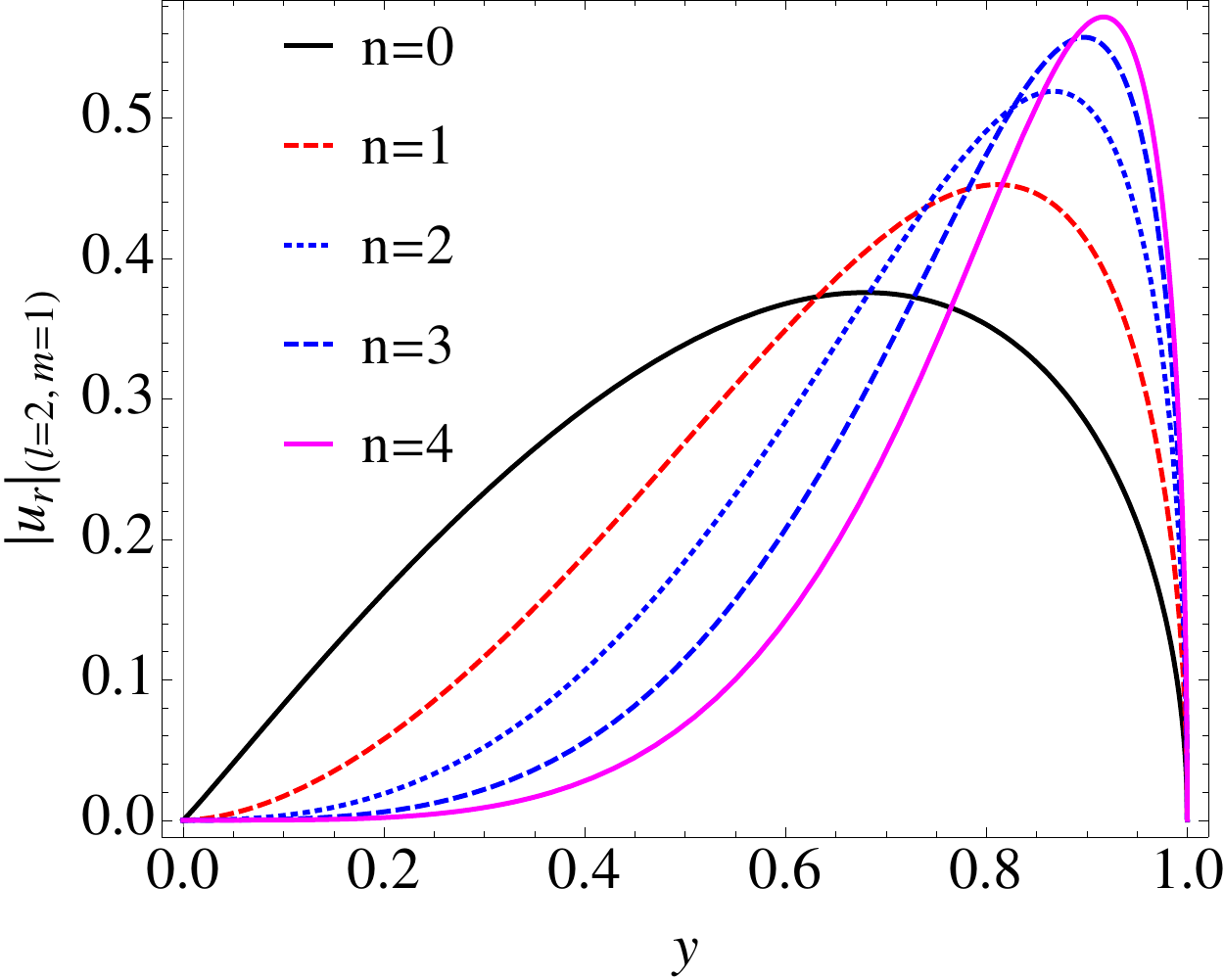}
\caption{ZDM wavefunctions in the bound-state picture for $(l,m,s)=(2,1,0)$. Black solid, red dashed, blue dashed, blue dotted and magenta solid lines correspond to $n=0,1,2,3,4$ respectively. The wavefunctions approach $0$ for $r \to r_+$ or $\sqrt{\epsilon} r_* \to 0$, and stay negligibly small for $r-1 \gg \sqrt{\epsilon}$. Therefore the wavefunction is bounded in the near-horizon regime. In addition, as we increase the overtone $n$, the wavefunctions monotonically move away from the horizon. This is a general feature for all ZDMs.
}
\label{fig:wavefunctionnearhorizon}
\end{figure}

We also examine the ZDMs in the dual bound-state picture. For simplicity, we focus on scalar perturbations $(s=0)$ and make the approximation that our transformation~\eqref{eq:transform} takes $A \to - A$\footnote{Actually, $A \to -A', \mathcal{F}^2_0 \to {\mathcal{F}^2_0}'$, and $A' \neq A$ for generic $s,l,m$. The value of $A'$ can be obtained using the expansion in~\cite{casals}. We make the approximation $A' =A$ for illustration (an estimate using the expansion gives $A' \approx 6.36$, while Leaver's method gives $A \approx 5.89$).}.
After the transformation, the near horizon radial equation~\eqref{eq:nearhorizon} becomes
\be
y^2 \frac{d^2 u}{d y^2}+y\frac{d u}{d y}-\left [ \tilde \omega^2+\frac{2\sqrt{2}m y}{1-y}\left (\tilde \omega-\frac{m}{\sqrt{2}}\right )+\frac{2 \mathcal{F}^2_0 y}{(1-y)^2}\right ]u=0 \,.
\ee
The solution can be written as a hypergeometric function similar to Eq.~\eqref{eq:hyper1},
\begin{align}\label{eq:hyper2}
u = y^{-p} (1-y)^{-q} {}_2 F_1 (\alpha,\beta,\gamma, y),
\end{align}
but with
\begin{align}
&p = - \tilde \omega/\sqrt{2}, &\qquad  q  = -1/2 -  \delta',  \notag \\
&\alpha = 1/2 +  (m + \delta'+  \sqrt{2} \bar \omega), & \qquad \beta = 1/2 - m +  \delta', \notag \\
& \gamma =  1 + \sqrt{2} \tilde \omega,
\end{align}
where $y=e^{\sqrt{2\epsilon}r_*}$ and $\delta' = \sqrt{\mathcal{F}^2_0-1/2}$. The resonant (bound state) condition is still $\gamma-\beta \approx -n$, or
\be
\tilde \omega \approx \frac{1}{\sqrt{2}}\left ( n+\frac{1}{2}+m-\delta' \right )\,
\ee
which transforms back to Eq.~\eqref{eq:PoleSln} if we apply the inverse transformation of Eq.~\eqref{eq:transform}. In Fig.~\ref{fig:wavefunctionnearhorizon} we plot the wavefunction for the first five overtones of the ZDMs with $(l,m)=(2,1)$. The wavefunction in the regime $r-1 \gg \sqrt{\epsilon}$ can be obtained using the matching method of Sec.~\ref{sec:matchex}. It is approximately zero, and it is not shown in the plot. One can similarly evaluate the DM wavefunctions in this dual picture, and see that their support is near the WKB peak, which is well separated from these ZDMs as $a \to 1$. Therefore this bound-state transformation provides us with a physical understanding of the two types of modes mentioned earlier: the DMs and ZDMs are each concentrated around different extrema of the radial potential, and this determines their behavior as we vary $\mu$ and $a$, according to the behavior and position of the extrema.

\subsection{Phase boundary}
\label{sec:PhaseTrans}

In the eikonal limit, the existence of a peak in the potential $-V_r$ outside the event horizon determines the critical value $\mu_c$ below which DMs and ZDMs coexist.
For general $l$ it is not clear that $\mu_c$ still defines the boundary between the two phases, but we expect that the peak of the potential should play the same role.

We provide a new analytic criterion for scalar perturbations of NEK, i.e. that there be no peak outside the horizon. We write the radial potential in the limit $a\to 1$ and assume that the frequencies are given by the ZDM branch only ($\omega = m/2$). This results in purely real ${}_0 A_{lm}$. Then $V_r$ is purely real, and given by
\begin{align}
V_r = &\frac{(r-1)^2}{(r^2+1)^2}\left [ \frac{(r+1)^2}{4}m^2-{}_0A_{lm}  \right ]\nonumber \\
&+\frac{(r-1)^2}{(r^2+1)^2}\left [\frac{3}{4}m^2+\frac{(r-1)(2r^2+3r-1)}{(1+r^2)^2} \right ] .
\end{align}
Solving for the roots of $V_r$ shows that the WKB condition extends to low $l$ and there is no peak outside the horizon when
\begin{align}
\mathcal F_0^2 = \frac{7m^2}{4} - {}_0 A_{lm} >0.
\end{align}

This criterion is more complicated for electromagnetic and gravitational QNMs. For these modes, the spin-dependent terms in $-V_r$ make the potential complex even in the extremal limit. It is not clear how to interpret the extrema of a complex potential in terms of our physical picture. However, Detweiler \cite{detweiler2} has shown that a suitably transformed new radial function satisfies a second-order differential equation with a real potential. This calculation is lengthy, but the result is a condition that there be no peak of the transformed potential outside the horizon, which holds for any $s$:
\be
\label{eqcriall}
\mathcal{F}^2_s\equiv \frac{7m^2}{4}-s(s+1)-{}_s A_{lm}\left (\omega=\frac{m}{2}\right )>0 .
\ee
The derivation of this condition is in Appendix~\ref{ap:DetTransform}. Note that this expression respects the pairing symmetry ${}_{-s}A_{lm}={}_sA_{lm}+2s$. 
Just as in the WKB case, the functions $\mathcal F_s^2$ and $\delta^2$ differ only by a factor of $1/4$, and we may ask whether the condition $\delta^2 >0$ differs from the condition~\eqref{eqcriall}. 
{There is also the even more important question as to whether the existence of a peak outside the horizon guarantees the existence of DMs.}

{If there is indeed a peak outside the horizon in the extremal Kerr limit, it
is often helpful to view the radial Teukolsky equation as a bound-state 
problem, as discussed in Sec.~\ref{sec:boundstate}. 
For a standard bound-state problem in quantum mechanics, no matter how shallow 
the potential well is, there is always at least one bound state. 
For generic Kerr BHs (except Schwarzschild BHs), however, the dual bound-state 
problem has more complicated dependence on the eigenvalue, and it is nontrivial
that any shallow potential well can support at least one bound state. 
In the original scattering problem, this means that even if there is a 
potential peak outside the horizon, there may not exist a DM associated with it.
As the potential well becomes deeper and wider ($\delta^2$ becomes more and 
more negative), we know from WKB analysis that DMs must exist for these peaks. 
In order to verify that DMs are associated with the presence of a peak outside the horizon, we need to go beyond our analytic (eikonal and near extremal) approximations.}

To answer these questions and to test the effectiveness of criteria \eqref{eqcriall}, we now carry out a series of numerical studies. 
These studies also illustrate how the QNM spectrum of Kerr changes as we enter the NEK regime.

\section{Spectrum Bifurcation}
\label{sec:Bifurcation}

In the previous sections, we have used analytical approximations to argue for the existence of a bifurcation in the Kerr QNM spectrum. In this section, we examine this bifurcation in greater detail numerically.

First of all, we verify that for the scalar ($s=0$) and gravitational ($s=-2$) perturbations with $2 \le l \le 100$, there are no QNMs which satisfy both $\delta^2 <0$ and $\mathcal F_s^2 >0$. 
This indicates that the sign of $\delta^2$ determines whether the potential has a peak outside the horizon. It is therefore not surprising that this seemingly arbitrary $\delta^2$ term would appear frequently in the analytical formulae.

For Schwarzschild and slowly spinning Kerr black holes, there is a single family of QNMs for given $(l,m)$ indexed by the overtone number $n$, and their decay rate monotonically increases with $n$. 
For fast-spinning Kerr black holes, however, there are two distinct families of QNMs with different decay properties, namely the DMs and the ZDMs. 
Therefore, there must be a transition at some large angular momentum where a single family of QNMs branches into two families with different properties. 
In this section we explore this spectrum bifurcation.

\subsection{Numerical investigation of the phase boundary}
\label{sec:NumPhase}

\begin{figure}
\includegraphics[width=0.48\textwidth]{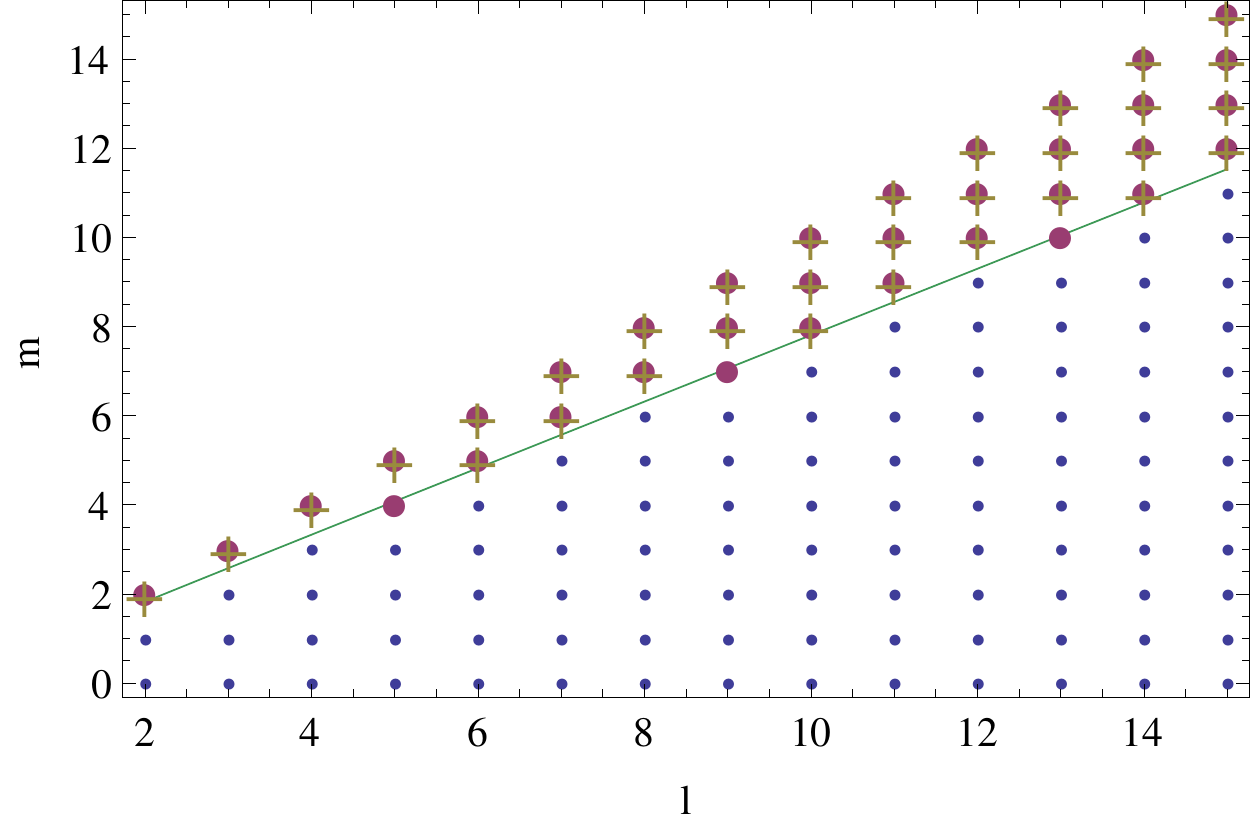}
\caption{Phase diagram for the separation between the single- and double-branch regime for NEK BHs. Large purple dots $(s=-2)$ and gold crosses $(s=0)$ correspond to $(l,m)$ pairs with only ZDMs, while smaller blue dots correspond to $(l,m)$ pairs with both ZDMs and DMs. The green line is the phase boundary, computed using the eikonal approximation. Reproduced from~\cite{Yang2012b}.}
\label{fig:phase}
\end{figure}

The interpretation of the DMs as unstable resonances associated with a potential peak outside the horizon is clear in the WKB limit, but we must test whether our analytic criteria hold outside of this limit. 
For this, we explicitly identified the phase boundary for $s = (0, -2)$,  $2 \leq l \leq 15$, and $0 \leq m \leq l$, by using Leaver's continued fraction algorithm, as discussed in Appendix~\ref{ap:NumericalMethods}. 
We searched for the presence of DMs at high angular momenta for $(l,\, m)$ values to each side of the the WKB prediction of the phase boundary for $s=(0,-2)$, and exhaustively checked for DMs below the phase boundary for $2 \leq l \leq 11, \, s = 0$. Combining our numerical exploration of the phase with our analytic understanding from Eq.~\eqref{eqcriall}, we show the $(l,\,m)$ pairs for which DMs exist in Fig.~\ref{fig:phase}, which is reproduced from~\cite{Yang2012b}. 
Small blue dots indicate those $(l,\, m)$ values where DMs are found along with the ZDMs at high angular momenta. Gold crosses indicate where only ZDMs exist for the scalar $(s=0)$ case, and large purple dots indicate where only ZDMs exist for the gravitational $(s=-2)$ case. The eikonal prediction for the phase boundary is shown as a green line.

The eikonal criterion $\mu < \mu_c$ for the existence of DMs is quite accurate even at low $l$. There are no scalar modes which violate this criterion. In the case of gravitational perturbations, three modes in this range of $l$ with values of $\mu$ very close to $\mu_c$ violate the bound. 
As discussed in~\cite{Yang2012a}, modes with nonzero spin have larger $O(L^{-2})$ errors relative to the WKB prediction than the scalar modes, so some discrepancy for $s\neq 0$ modes is expected. 

This numerical investigation verifies our understanding that the DMs are associated with a peak in the potential outside the horizon. 
With our physical picture confirmed, we turn to the problem of connecting the two-phase nature of the QNM spectrum with the usual understanding of a hierarchy of QNMs indexed by an overtone number $n$.

\subsection{Numerical investigation of bifurcation}
\label{sec:NumBifurcation}

In this section, we explore the bifurcation of the QNM spectrum as the angular momentum of the black hole increases. We focus on gravitational, primarily quadrupolar, corotating QNMs $(s = -2, \, l= 2, \, m \geq 2)$. 
These modes have astrophysical relevance, especially the $l = 2, \, m=2$ mode, which dominates the ringdown following black hole mergers \cite{Buonanno:2006ui,Berti:2007fi}. 
Numerical exploration of this case is especially important, since the spectrum bifurcates at a relatively small value of the angular momentum of the hole, and also the WKB approximation is worst for these small $l$, gravitational modes. 
The case of $s=0,\ l= 10$ modes near the phase boundary was already discussed in~\cite{Yang2012b}, and so we only briefly discuss the bifurcation of the $s=0,\ l=10,\, m=7$ mode here. 
For these large-$l$ modes, the WKB and NEK approximations give an accurate description of the system (although the WKB formalism in~\cite{Yang2012a} treats only the lowest-overtone DM accurately). 
Finally, we develop an estimate for the critical angular momentum at which the bifucation occurs for a given mode, using our analytic tools.

We use two methods to explore the QNM spectrum of the NEK spacetime. 
Both rely on the standard continued-fraction technique for finding QNM frequencies introduced by Leaver~\cite{leaver}. 
In this technique, the QNM frequencies are the zeroes of a certain infinite continued-fraction expansion, which in practice is truncated at some high order in the expansion. 
Our first method is to search for these roots of the continued fraction directly, as we vary the angular momentum of the black hole. 
This allows us to track the behavior of a single QNM as the bifurcation occurs. 
The second method is to directly compute the value of the continued fraction throughout some region of the complex plane, and plot contours of constant value of the fraction. 
Those places where the value is approximately zero identify the QNMs. 
This allows us to identify many overtones simultaneously for a chosen set $(s,\,l,\,m)$ and a fixed angular momentum. 
We discuss our numerical methods further in Appendix~\ref{ap:NumericalMethods}.

\subsubsection{The $s=-2,\,l=2$ modes}
\label{sec:L2GravModes}

\begin{figure*}[t]
\includegraphics[width=2.0\columnwidth]{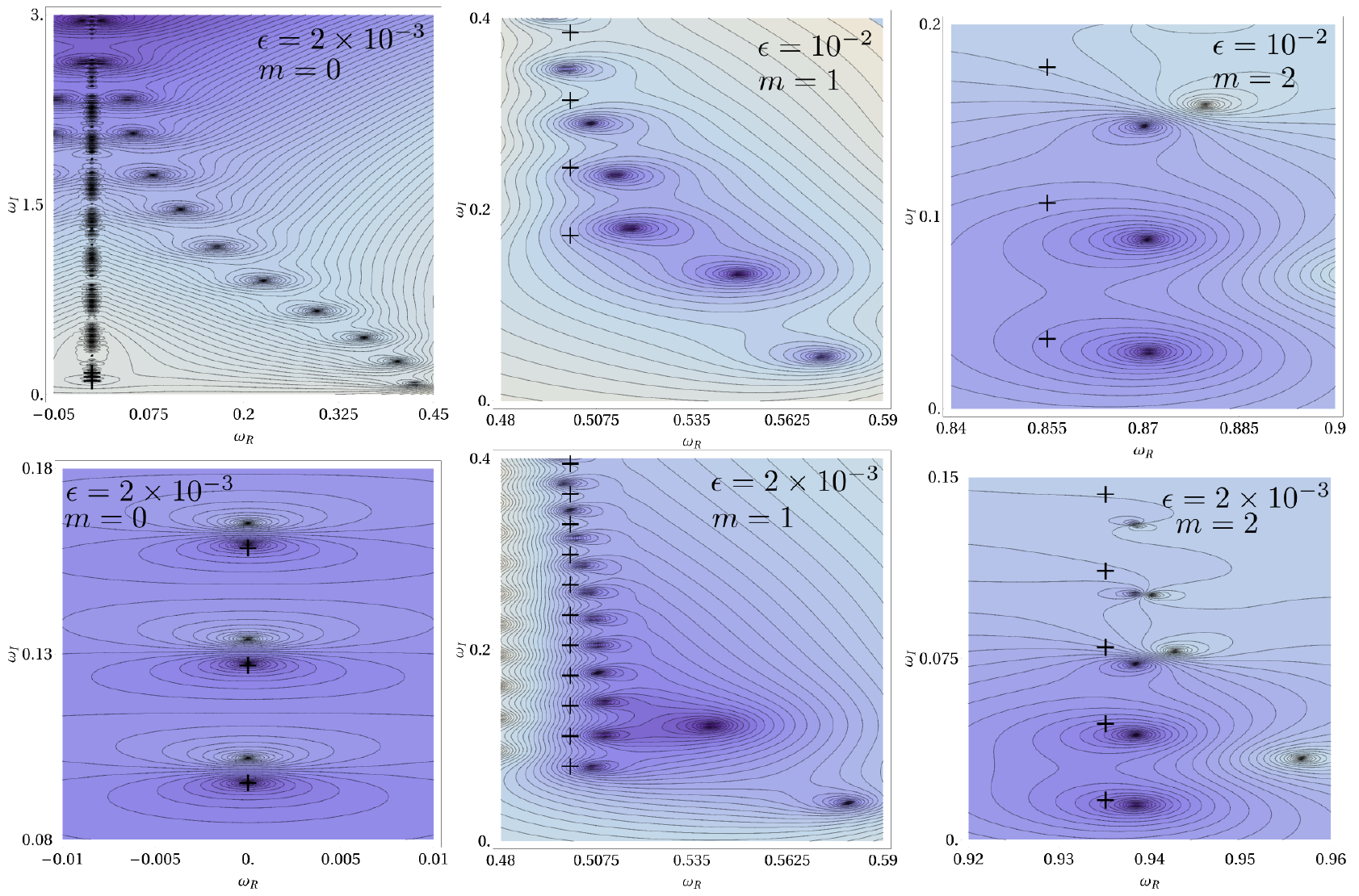}
\caption{Contours are constant values of the logarithm of the continued fraction  for the case $s = -2$, $l=2$, $m\leq 2$.Darker shading indicates values close to zero. The contours cluster around the QNM values in darker regions, and around poles in the fraction for lighter regions. The $+$'s correspond to the analytic ZDM prediction, Eq.~\eqref{eqhod2}.}
\label{fig:L2Contours}
\end{figure*}

For $s=-2$, $l=2$, we examine $m= 0,\, 1$ and  $2$ in turn. Each of these values of $m$ corresponds to a different phenomenology of the spectrum when ZDMs are present. Figure~\ref{fig:L2Contours} illustrates the behavior of the QNMs for the NEK spacetime in these cases by plotting the logarithm of the value of the continued-fraction expansion in a region of the complex plane. The darker color indicates values close to zero, and the contours cluster around the QNMs as well as around poles of the continued fraction. These poles are often paired with a QNM and can be distinguished by their relatively lighter shading. They have no physical significance, and their position is changed by inverting the continued fraction expansion at various places, while the QNMs remain unchanged.

The case $m=0$ is marginal, in the sense that there are no predicted ZDMs for any $m<0$. 
For $l=2$, we find that there are a large number of DMs with $\omega_R \neq 0$ at high angular momenta, and a very large number of QNMs on the imaginary axis. 
Some of these modes do not change in decay rate according to Eq.~\eqref{eqhod2}, and so these are DMs, but it is not clear if there is a finite number of these modes. 
Meanwhile, the ZDMs are also present, with oscillation frequencies and decay rates that are in agreement with Eq.~\eqref{eqhod2}. 
The top left panel of Fig.~\ref{fig:L2Contours} illustrates for $a = 0.998$ the eleven DMs with $\omega_R \neq 0$, as well as a twelfth mode on the imaginary axis which does not decrease in decay as the angular momentum is increased. 
Some of the negative-frequency modes are also visible. 
The ZDMs can be seen as small, dense clusters of contours. 
The bottom-left panel of Fig.~\ref{fig:L2Contours} zooms in on the first three ZDMs, showing that they are accompanied by poles and are well described by the analytic approximation. 
Since the angular momentum at which the first ZDM has a decay rate less than the eleventh DM must be quite low ($a \lesssim 0.16$ by our NEK expressions, which are not valid for such small values of $a$), we can only estimate at what angular momentum the spectrum bifurcates.

The case $m=1$ is close to the phase boundary, and provides a clean example of the bifurcation of the spectrum. 
The top, middle panel of Fig.~\ref{fig:L2Contours} shows the first six QNMs for $a = 0.99$. 
They have monotonically increasing decay and can be labeled with a single overtone index. 
As the angular momentum increases to $a = 0.998$, the spectrum bifurcates, as seen in the bottom, middle panel of Fig.~\ref{fig:L2Contours}. 
There are two DMs in this case, and as the spectrum divides the overtone $n$ QNMs become the $n' = n-2$ ZDMs for $n>2$. 
Though this bifurcation behavior occurs for the NEK, the angular momentum at which the branches separate in the $(2,1)$ case is actually not beyond the scope of what may be achieved by astrophysical black holes~\cite{Thorne}. 

\begin{figure}[t!]
\includegraphics[width=1.0\columnwidth]{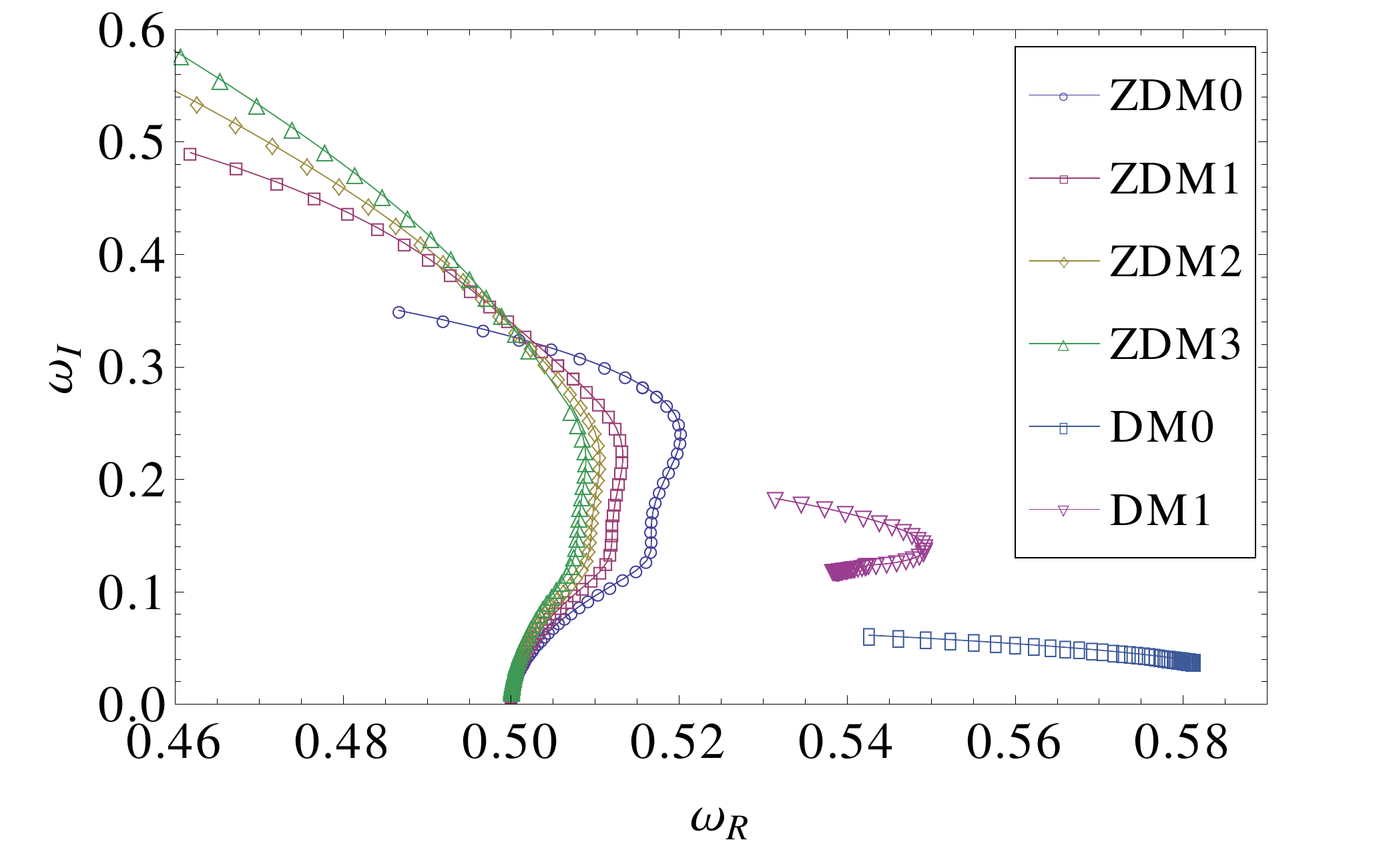}
\caption{Plots of QNM frequencies $\omega$ for the case $s = -2$, $(l,\,m) = (2,1)$, values as found using Leaver's method (with inversion). The first six overtones (at low angular momenta) are shown, which become two DMs and the first four ZDMs as we increase $a$ from $0.9$ to $0.99999$, using logarithmically decreasing spacing. The decay rate $\omega_I$ of the two DMs changes relatively little as $a\to 1$, while the ZDMs move towards their extremal limit $\omega \to m/2$.}
\label{fig:Modes21sm2}
\end{figure}

In order to further illustrate the behavior of the bifurcation for the $(2,1)$ mode, in Fig.~\ref{fig:Modes21sm2} we plot the first six overtones (at low angular momentum) as we increase the angular momentum from $a= 0.9$ towards the extremal value, ending at $a=0.9999$. 
The two DMs change relatively little, while the decay rates of the first four ZDMs begins to rapidly decay after these modes cross the vertical line $\omega_R = m/2$. 
In fact, all four modes pass through nearly the same frequency value (although at different angular momenta), $\omega = m/2 - 0.325 i$. 
The nature of this ``focusing'' frequency is not clear, but it appears to mark the onset of the NEK regime for each mode. 
The results of Fig.~\ref{fig:Modes21sm2} match those found by Leaver, in Fig.~3 of~\cite{leaver} (although note Leaver's convention $M = 1/2$).

Finally, $m=2$ features only ZDMs and no DMs. We show in the top-right and bottom-right panels of Fig.~\ref{fig:L2Contours} the first few QNMs for two values of the angular momentum. As we increase the angular momentum from $a = 0.99$ to $a = 0.998$ we see that the ZDMs change, as expected from Eq.~\eqref{eqhod2}, approaching their final values at $\omega = m/2$. The NEK prediction becomes more accurate with increasing spin. We note that there are several poles evident in the right panels of Fig.~\ref{fig:L2Contours}, which are distinguished by their lighter shading, and again emphasize that these have no physical relevance.

We present a discussion of the scalar $l=2,\, m=1$ QNMs in Appendix~\ref{sec:ExtraMode}. These modes behave similarly to the gravitational modes, but exhibit some curious behavior, which may serve as the subject of future study.

\subsubsection{The $l=10$ modes}

\begin{figure}
\includegraphics[width=1.0\columnwidth]{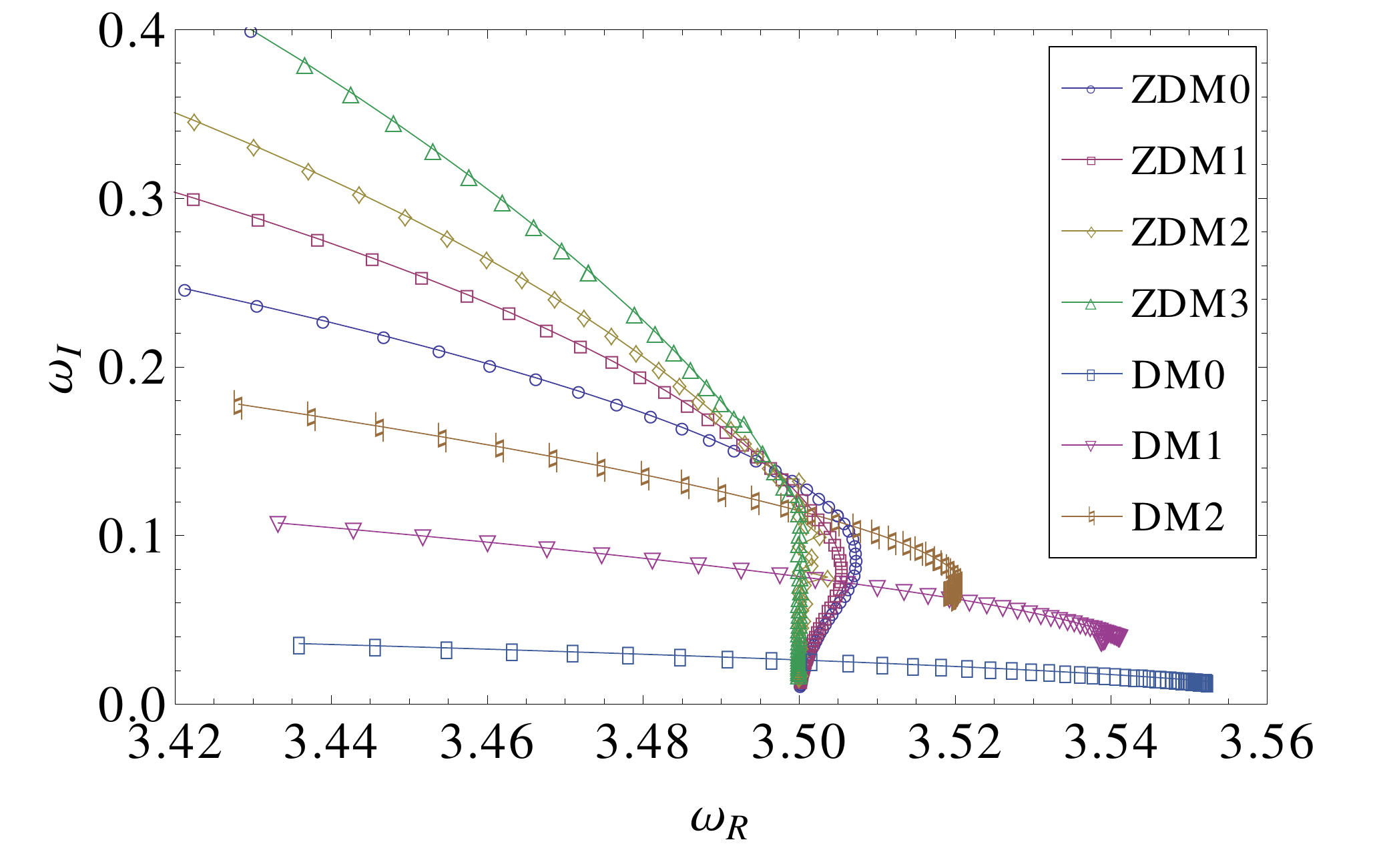}
\caption{Plots of QNM frequencies $\omega$ for the case $s = 0$, $(l,\,m) = (10,7)$, values as found using Leaver's method (with inversion). The first seven overtones (at low angular momenta) are shown, which become three DMs and the first four ZDMs as we increase $a$ from $0.99$ to $0.99999$, using logarithmically decreasing spacing.}
\label{fig:ModesL10M7}
\end{figure}

Contours showing the spectrum bifurcation in the case $l=10$ are presented in \cite{Yang2012b}. Here we briefly supplement those results by plotting the trajectories of the QNM frequencies as the angular momentum of the black hole increases for $m=7$.
Figure~\ref{fig:ModesL10M7} plots the trajectories of the first seven overtones as the hole's angular momentum increases, in the same manner as Fig.~\ref{fig:Modes21sm2}. 
Comparing to the frequency trajectories for the gravitational $(2,1)$ mode, we see the same general behavior for the QNM overtones: initially, for angular momentum $a = 0.99$, the modes have monotonically increasing decay rates. As the angular momentum increases, the frequency of the first three overtones increases as their decay rate remains nearly constant. 
These become the three DMs following bifurcation. The higher overtones approach and pass through approximately the same frequency $\omega = m/2 - 0.15i$ (although at a different value of $a$ for each overtone), after which their decay rates decrease rapidly to values below those of the three DMs. 
These are the ZDMs. We note that at high angular momenta, many ZDMs exist in close proximity to the line $\omega_R = m/2$, and in this case we have difficulty resolving the paths of the third (yellow diamonds), fourth (green triangles), and higher ZDMs after they pass through the focusing frequency.

\subsubsection{Critical angular momentum for ZDMs}

Our results so far raise the question: for a given $(s,\,l,\,m)$, what is the critical value of the angular momentum $a$ where the bifurcation begins? This question could be answered by an exhaustive numerical investigation. A much faster method gives an upper bound on the angular momentum at which bifurcation occurs, in the following way. We ask for which angular momentum $a_c$ the least-damped ZDM reaches the same decay rate as the least-damped DM; of course, bifurcation would be better said to occur when the least-damped ZDM has the same decay as the most-damped DM, since this is where the two branches become distinct. Nevertheless, the angular momentum $a_c$ can be estimated fairly easily using our WKB results for the least-damped DM. Setting the two decay rates from Eqs.~\eqref{eq:OmegaI} and~\eqref{eqhod2} equal at $\epsilon_c = 1 - a_c$, we have the equation
\be
\label{eqcrita}
\frac{\sqrt{\epsilon_c}}{2\sqrt{2}}(1+2|\delta|)=\left. \frac{1}{2}\frac{\sqrt{2V''_r}}{\partial_{\omega}V_r}\right |_{r_0}.
\ee
We solve this iteratively, since both sides depend on $\epsilon_c$, and the iterative solution converges rapidly for a variety of initial guesses for $\epsilon_c$.

\begin{figure}[t!]
\includegraphics[width=1.0\columnwidth]{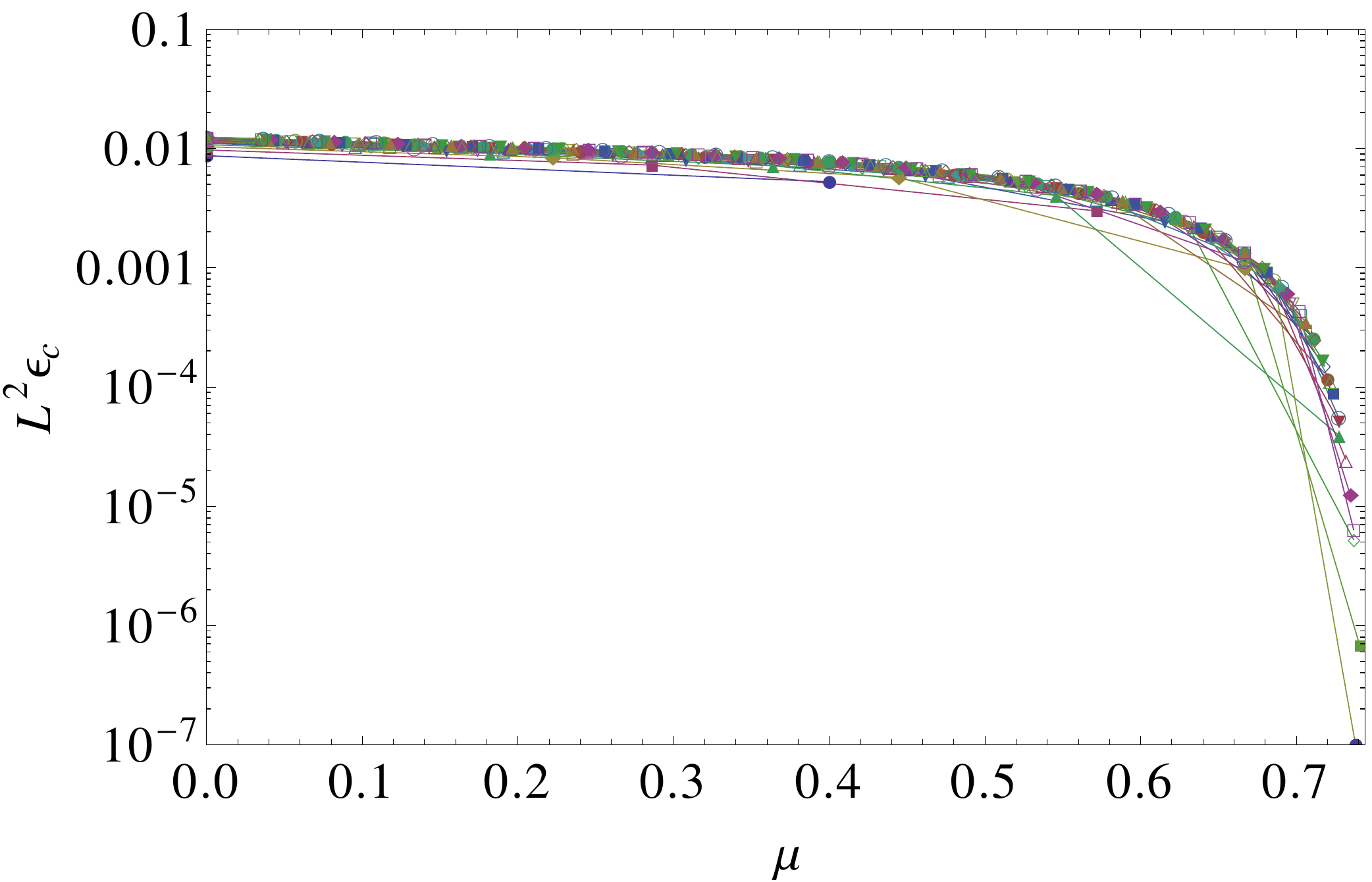}
\caption{Estimates for the (scaled) critical values $L^2 \epsilon_c$ for which $\mathcal{I}m (\omega_{\rm ZDM}) = \mathcal{I}m (\omega_{\rm DM})$, when the lowest order ZDM becomes the least-damped QNM. All estimates for modes with $2\leq l \leq 15$ and $0 \leq m < (l + 1/2) \mu_c$ are plotted, with lines linking estimates with the same index $l$. The scaled values cluster onto a curve which depends only on $\mu$. }
\label{fig:BifurcationSpins}
\end{figure}

In Fig.~\ref{fig:BifurcationSpins} we plot $L^2 \epsilon_c$ as computed using the iterative method; we see that, once scaled by $L^2$, the estimated $\epsilon_c$ values cluster tightly onto a limiting curve. A quadratic fit to this curve provides the reasonably accurate formula 
\begin{align}
\label{eq:CriticalLine}
L^2 \epsilon_c = 10^{-3}(11.7 - 3.39 \mu - 17.7 \mu^2).
\end{align}
These fitting coefficients differ slightly from those presented in \cite{Yang2012b}, because here the $m=0$ values have been included in the fit. The formula~\eqref{eq:CriticalLine} evaluated at $(10,7)$ gives $\epsilon_c = 1.44\times10^{-5}$, while numerically we can estimate $\epsilon_c \approx 1.5 \times 10^{-5}$. The simple limit curve suggests that there may be a simple analytic expression for $\epsilon_c$ in the eikonal limit. But the computation of the angular momentum at which the branches actually bifurcate remains an open question. The larger the number of DMs, the worse the estimate $\epsilon_c$ is. From our investigation of the $m=0$ cases, we see that there can be quite a large number of DMs for smaller values of $\mu$, and in these cases $\epsilon_c$ is a poor estimate. 

\section{Power-law ringdown of NEK excitations}
\label{sec:Decay}

In this section, we turn to a different application of the NEK frequency formula~\eqref{eqhod2}. The slow fall-off of the NEK ZDMs suggests that the ringdown of rapidly rotating black holes might be a promising signal for gravitational wave observatories. However, the high quality factor of the oscillations means that these modes are only weakly excited by any perturbing source \cite{Berti:2006wq,Zhang:2013ksa}, and therefore have prohibitively small amplitudes. 

The weak excitation factor of each mode can be overcome by exciting many weakly damped overtones together. As argued by Glampedakis and Anderson~\cite{Glampedakis:2001js}, this collective excitation can result in a perturbation which decays as $1/t$. Though in~\cite{Glampedakis:2001js} the $1/t$ decay was shown to set in at late times, we find that the power-law ringing occurs at \textit{early} times (and is distinct from the familiar late-time power-law tail).  Essentially, these weakly damped overtones all share very similar $\omega_R$, and combine into a super-mode. Their $\omega_I$'s are nearly evenly spaced (see Eq.~\eqref{eqhod2} and Fig.~\ref{fig:L2Contours}), so the super-mode's decay rate contains a multiplicative factor given approximately by
\begin{align}
\label{eq:TailSimplification}
\sum_{n=0}^\infty e^{-(n+1/2) \sqrt{\epsilon/2} t} &=  \frac{e^{-\sqrt{\epsilon/8} t}}{1-e^{-\sqrt{\epsilon/2} t}} \notag \\ & \approx \frac{1}{\sqrt{\epsilon/2} t}  + \mathcal{O}\left(\sqrt{\epsilon/2} t\right) \,,
\end{align}
which is dominated by the $1/t$ term for small $\sqrt{\epsilon/2} t$.
In~\cite{Glampedakis:2001js}, analytic and numerical computations are given that demonstrate this power-law decay, but inaccurate QNM frequencies are used to derive the analytic results. Cardoso argued in~\cite{cardoso} that only a relatively small number of modes contribute to the excitation, again using an inaccurate frequency formula. It is therefore interesting to reconsider this problem using our current understanding of the NEK QNM spectrum. In Sec.~\ref{sec:DecayCalc} we show that a generic perturbation excites many overtones resulting in a super-mode, which initially decays as $1/t$. In Sec.~\ref{sec:NumDecay} we provide numerical evidence for the power-law decay of NEK perturbations.

\subsection{Calculation of the polynomial ringdown}
\label{sec:DecayCalc}

Following the calculation of ~\cite{Glampedakis:2001js}, we consider a generic scalar perturbation of a NEK spacetime at an initial Boyer-Lindquist time slice. Consider initial data for the scalar wave equation $\Phi_0 = \Phi(t = 0, r, \theta,\phi) $ and $\dot \Phi_0 = \partial_t \Phi(t = 0, r, \theta,\phi)$. These initial data correspond to a source term $\mathcal S(r,\theta,\phi)$ which we integrate over using the Green's function approach. The source term is defined as\footnote{Note that in~\cite{Glampedakis:2001js}, the source term is already expanded in azimuthal harmonics.}
\begin{align}
\label{eq:Source}
\mathcal S =& \frac{\sqrt{r^2 +a ^2}}{\Delta}\biggl( \left [ i \omega (r^2 +a^2)^2 - 4 i a m r - i \omega a^2 \Delta \sin^2 \theta \right]  \Phi_0 \notag \\ 
& - \left[ (r^2+a^2)^2 - a^2 \sin^2 \theta \right] \dot \Phi_0 \biggr) \,.
\end{align}
We expand this source term using the spheroidal harmonics,
\begin{align}
\mathcal S_{lm} = \int \sin \theta d \theta d \phi \, S_{lm}^*(\theta) e^{- i m \phi} \mathcal S(r, \theta, \phi)\,.
\end{align}
By expanding $\Phi(x)$ in the frequency domain, we can write the scalar field at times $t > r_* + r_*'$ by integrating the source term using the radial Green's function $G(r_*, r_*')$. The result is~\cite{Glampedakis:2001js}
\begin{align}
\Phi(x) =& \sum_{l,|m|\leq l} \frac{\Phi_{lm}(t,r,\theta)}{\sqrt{r^2+a^2}} e^{i m \phi} \\ 
\label{eq:GreenInt}
\Phi_{lm}  =&  \frac{1}{2 \pi  }\int d \omega dr'  e^{- i \omega t} S_{lm}(\omega, \theta) \frac{ G (r,r') \mathcal S_{lm} (\omega, r')}{(r')^2 +a^2}   \,.
\end{align}
Although we mostly use the notation of~\cite{Glampedakis:2001js} in this section, note that $\Phi_{lm} = {}_0 u_{lm}$, recalling that ${}_0 u_{lm}$ is the radial function used in the radial Teukolsky equation~\eqref{eqgeneralteuk}, with $s=0$. 
For simplicity, we assume that the source is localized at a large radius and that the observer is also at a large radius. This reduces the Green's function to the simple form
\begin{align}
\label{eq:GreenFunc}
G(r,r') = -\frac{e^{i \omega r_*}}{2i\omega}\left( \frac{A^{\rm out}_{l m n}}{A^{\rm in}_{l m n}} e^{i \omega r_*'} + e^{-i \omega r_*'}\right)\,,
\end{align} 
The terms $A^{\rm out}$ and $A^{\rm in}$ are the asymptotic amplitudes for the ingoing and outgoing waves in the ``in'' wave solution, which represents a solution for ${}_0 u_{lm}$ that is purely ingoing at the horizon,
\begin{equation}
u^{\rm in} \sim \left\{
\begin{array}{cl}
e^{-i (\omega - m \Omega_H) r_*}\,, & r \rightarrow r_+\, \\
\\
A^{\rm out}e^{i \omega r_*}+A^{\rm in}e^{-i \omega r_*}\,, & r \rightarrow +\infty\,
\end{array}\right. \,.
\label{ax}
\end{equation}
The zeros of $A^{\rm in}(\omega)$ correspond to the QNM frequencies for which the waves are purely outgoing at spatial infinity and ingoing at the horizon. These correspond to simple poles of the Green's function~\eqref{eq:GreenFunc}.
Focusing on the QNM contribution to the scalar field harmonics $\Phi_{lm}$, we deform the contour of integration over $\omega$ in Eq.~\eqref{eq:GreenInt} into the lower half plane, and the integral is converted into a sum over the residues of the poles $\omega_{lmn}$. In this case only the first term in the parenthesis of the Green's function~\eqref{eq:GreenFunc} contributes, and the contour integral resolves as
\begin{align}
\label{eq:Phit}
\Phi_{lm} \approx &-\frac{i}{2}\int d r' \sum_n \frac{A^{\rm out}_{lmn}}{\alpha_{lmn}} e^{-i\omega_{lmn} T} \frac{\mathcal S_{lm} (r')}{i \omega_{lmn} (r')^2}  S_{lm}
\end{align}
where $\alpha_{lmn} =d A^{\rm in}/d\omega\, |_{\omega=\omega_{lmn}}$, and we have defined $T = t - r_* - r'_*$.
For the ZDMs, expressions for $A^{\rm in}$ and $A^{\rm out}$ are given in Appendix~\ref{ap:NEKCalcs}. The important point for the current discussion is that, keeping only the leading order terms in $\epsilon\ll 1$ and $\eta \ll 1$ and using Eq.~\eqref{eq:PoleSln}, we can write
\begin{align}
\label{eq:ExFactors}
\left . -\frac{i}{2} \frac{ A^{\rm out}}{\alpha}\right |_{ZDM}& \approx C(m,\delta) \frac{e^{- i\delta \ln 8\epsilon + i n \pi} \,\sqrt{\epsilon}}{n! \Gamma[-n +2 i \delta]} \,,
\end{align}
where the constant $C(m,\delta)$ is only weakly dependent on the overtone number $n$, through higher-order terms in $\epsilon$. Inserting this into Eq.~\eqref{eq:Phit} allows us to perform the sum over overtones explicitly,
\begin{align}
\Phi_{lm} & \approx  C e^{-i \delta \ln 8 \epsilon}   S_{lm} \int dr'  \sqrt{\epsilon}\,
e^{- i m T/2 -\sqrt{\epsilon/8} T} \notag \\ & \qquad \qquad \times
\sum_n \frac{e^{-n\sqrt{\epsilon/2} T+ i n\pi}}{n! \Gamma[-n +2 i \delta]}  \frac{2 \mathcal S_{lm}}{i m (r')^2}
\\
\label{eq:TailResponse}
& \approx \frac{C e^{-i \delta \ln 8 \epsilon}}{\Gamma[2 i \delta]} S_{lm} \int dr'  \frac{
 \sqrt{\epsilon}\,  e^{- i m T/2 -\sqrt{\epsilon/8} T}}{1- e^{- \sqrt{\epsilon/2}T}} 
\notag \\ &  \qquad  \qquad \times
\left(1 - e^{-\sqrt{\epsilon/2} T} \right)^{2i \delta}
 \frac{2 \mathcal S_{lm}}{i m (r')^2} \,.
\end{align}
For early times, $T >0$ and $ T \sqrt{\epsilon/2} \ll 1$, the integrand in Eq.~\eqref{eq:TailResponse} reduces to
\begin{align}
\label{eq:Integrand}
 \frac{\sqrt{2}e^{-imT/2}}{T}  \left(\sqrt{\epsilon/2}\, T\right)^{2 i \delta} \frac{2 \mathcal S_{lm}}{im(r')^2} \,.
\end{align}
For a fixed $r_*$, this gives the $\sim 1/t$ dependence of the amplitude of the QNM ringing, provided $\delta^2 >0$, so that $\delta$ is purely real and contributes only to the phase of the integrand~\eqref{eq:Integrand}. 
We see that the power-law decay is only valid for about $1/\sqrt{\epsilon}$. 
At later times, the power-law falloff transitions to an exponential decay dominated by the $n=0$ overtone. 
This explains recent results by Harms et al.~\cite{Harms:2013ib}, whose numerical investigations of perturbed, rapidly rotating Kerr holes shows exponential decay at late times unless $\epsilon \to 0$. 
In this case, the limit of the integrand gives a decay for the response function equal to $\sqrt{2}/T$. 
A careful examination of Fig.~17 of~\cite{Harms:2013ib} shows that for large angular momenta at early times ($T \sqrt{\epsilon/2} < 1$), the asymptotic perturbation decays slower than exponential, and in fact seems to obey a roughly $1/T$ dependence (that figure actually plots the gravitational, $s=-2$ ringdown, but the spin dependence enters only in the coefficient $C$). We build on these results with our own numerical experiments in the next section.

\begin{figure}
\includegraphics[width=1.0 \columnwidth]{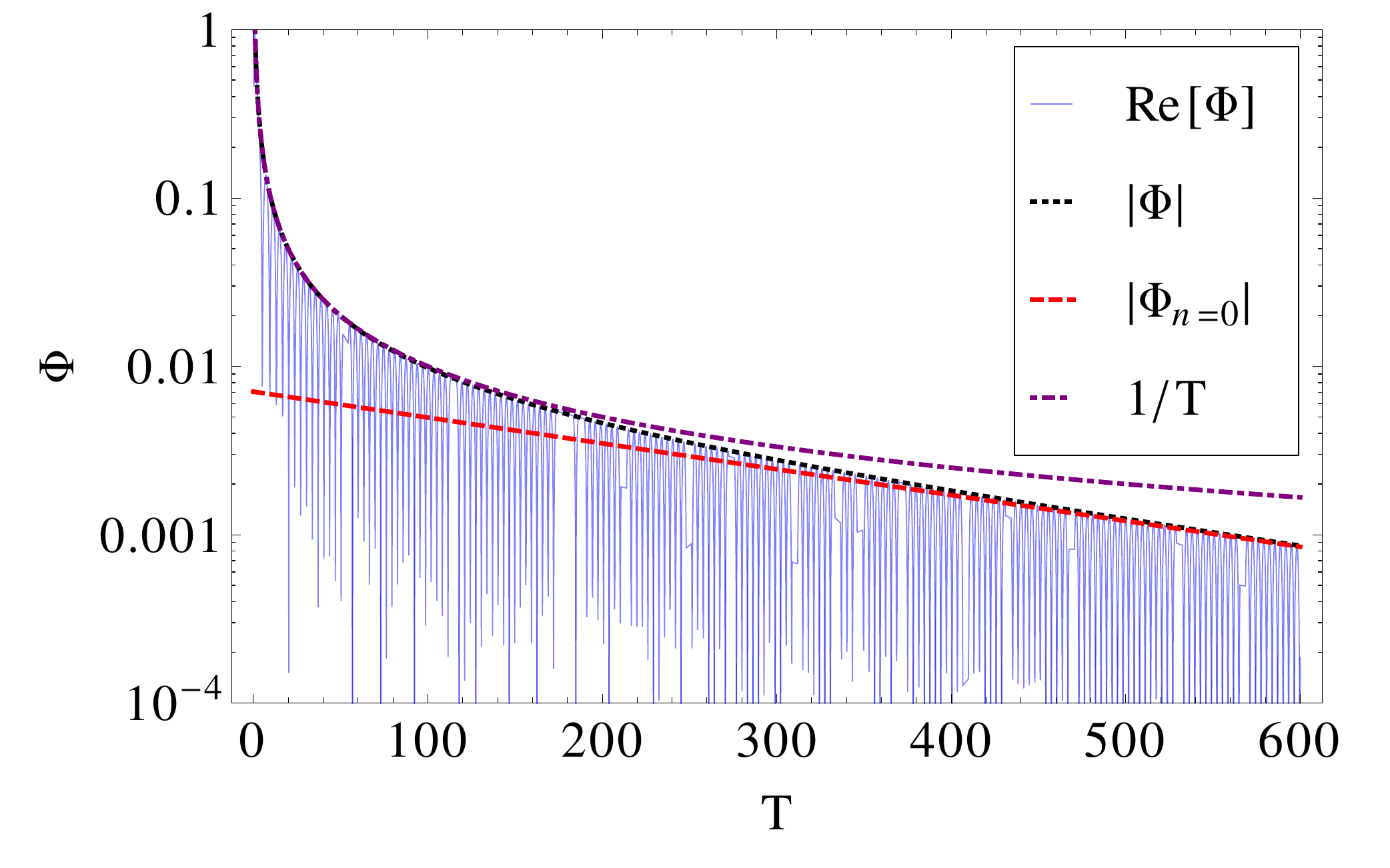}
\caption{The scalar QNM response of a hole with $\epsilon = 10^{-4}$, as described by Eq.~\eqref{eq:TailResponse} for a scalar $l=2,\,m=2$ perturbation, with a radial $\delta$-function distribution. The initial amplitude is normalized to unity. The blue curve plots the logarithm of $\Re[\Phi](T)$, and its envelope is shown by a black dotted line. Also plotted is a curve $1/T$ (purple dot-dashed line) and the decay envelope for the $n=0$ ZDM (red dashed line).}
\label{fig:Tail}
\end{figure}

As an example, we take for our initial data $\dot \Phi_0 = 0$ and let $\Phi_m(0,r,\theta) = A \delta(r - r_0) S_{22}(\theta) \delta_{m2}$. This simplifies the source term of Eq.~\eqref{eq:Source} to
\begin{align}
\mathcal S_{lm} = i \omega_{lmn} A r^3 \delta(r-r_0) \delta_{m2} \,,
\end{align}
in the limit that $r_0 \gg a$, the asymptotic scalar field at $r \to \infty$ reads
\begin{align}
\label{eq:asympt}
\Phi(x) \approx & A \frac{r_0}{r} S_{22}(\theta) e^{2i\phi}   \frac{ C' e^{i \delta \ln 8 \epsilon}}{(1- e^{-\sqrt{\epsilon/2} \, T})^{- 2 i \delta}}  \notag \\ & \times
\frac{ \sqrt{\epsilon} e^{- i  T-\sqrt{\epsilon/8} T}}{1- e^{- \sqrt{\epsilon/2}T}} \,.
\end{align}
This expression agrees with our expectation from Eq.~\eqref{eq:TailSimplification}. In Fig.~\ref{fig:Tail} we illustrate the time dependence of this QNM ringing by plotting the time dependent amplitude of $\Re [\Phi]$ at some fixed radius and angle, with the amplitude normalized to unity at $T = t- r_* -r_{0*} = 0$, so that at this time the functional behavior limits to $1/T$. 

In the case where there are also DMs, we can evaluate their contribution by including additional exponentially decaying contributions to $\Phi_m$. A key observation for these modes is that $\delta^2 <0$ when DMs are present. Taking the convention $\delta = i |\delta|$, the analysis of the ZDM response carries forward directly, though phase factors in the previous case can contribute to the amplitude when $\delta$ is imaginary. Aside from an adjustment in phase and amplitude, the decay of the mode at early times is modified, since the integrand~\eqref{eq:Integrand} times the factor $e^{|\delta| \ln{8 \epsilon}}$ becomes
\begin{align}
e^{-imT/2} 2^{|\delta|+1/2}   T^{-1-2 |\delta|} \frac{2 \mathcal S_{lm}}{im} 
\end{align}
in this case. We see that the early-time decay is stronger than when only the ZDMs are present. At late times, we once again recover the exponential decay of the least-damped QNM, although there is an additional suppression of the late-time amplitude by a factor of $\epsilon^{|\delta|}$. Additionally, in this case the ringing of the DMs competes at early times with the power-law decay and may dominate, but at later times the least-damped ZDM must still dominate the ringing.

\subsection{Numerical study of polynomial ringdown}
\label{sec:NumDecay}
The slow ringdown of collective ZDMs calculated in the previous section relies on the near-extremal limit. To test the results with minimal assumptions we perform a numerical simulation without any approximations other than discretization. We solve the scalar wave equation on a Kerr background with various angular momenta. The parameter $\epsilon$ ranges from $10^{-2}$ to $10^{-6}$ in our simulations, which gives us a wide enough range to study the transition from the generic exponential ringdown to the slow, polynomial ringdown. 

We solve the homogeneous scalar wave equation on a Kerr background 
\be\label{eq:wave} \Box \Phi = 0,\ee
in symmetric hyperbolic, fully first order form with auxiliary variables 
$\Phi_i:= \partial_i \Psi$ and $\Pi:=-1/\alpha(\partial_\tau \Psi-\beta^i \partial_i \Psi)$, as in \cite{Scheel:2003vs}. Here, $\alpha$ is the lapse and $\beta^i$ is the shift. We study the evolution of the $l=2,m=2$ mode, so initial data are given as 
\begin{align} \label{eq:id} \Phi(0,{\bf x}) &= 0, \ \Phi_i(0,{\bf x}) = 0, \\
\Pi(0,{\bf x}) &= e^{-(r-r_0)^2/\sigma^2}  \Re[Y_{22}(\theta,\varphi)], \end{align}
where ${\bf x}=\{r,\theta,\varphi\}$ are the spatial coordinates. We set $r_0=1.2$ and $\sigma=0.2$. The initial Gaussian is always centered outside (but close to) the horizon.

For our numerical simulations we use the spectral Einstein code (SpEC), which is a spectral element code for solving elliptic and hyperbolic partial differential equations \cite{SpECWebsite}. A spectral expansion in space is performed in elements that communicate with each other along touching internal boundaries through the exchange of characteristics via penalty terms. We use Chebyshev polynomials with Gauss--Lobatto collocation points in the radial direction and a spherical harmonic expansion in the angular direction. The discretized unknown is then evolved as a coupled system of ordinary differential equations in time with adaptive time-stepping using the Dormand--Prince method. 

We employ explicit, scri-fixing, hyperboloidal compactification \cite{Zenginoglu:2007jw} to compute the unbounded domain solution and the signal at infinity as measured by idealized observers. Computational efficiency is crucial for simulations on fast-rotating black-hole backgrounds. To achieve high efficiency, we use smooth, horizon-penetrating, hyperboloidal coordinates as in recent time-domain simulations in Kerr spacetime \cite{Racz:2011qu, Harms:2013ib}. 
For regularity of the wave equation \eqref{eq:wave} at null infinity, we solve the equation for a rescaled field that asymptotically corresponds to $r \Phi$. 

\begin{figure}
\includegraphics[width=\columnwidth]{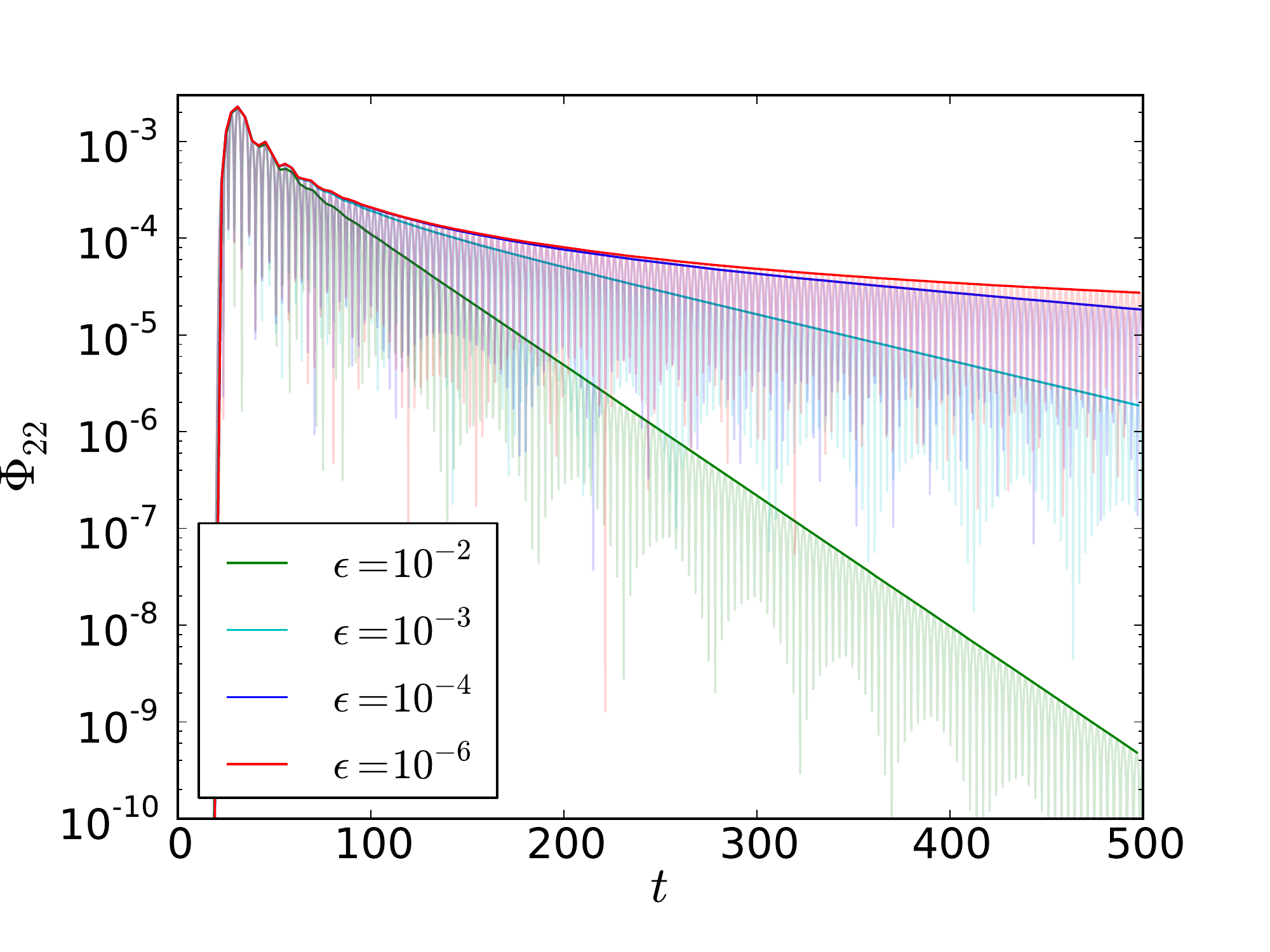}
\caption{Time domain signal for the $l=2,m=2$ mode measured at future null infinity for various angular momenta. Early-time polynomial ringdown is observed longer for smaller $\epsilon\equiv 1-a$ in accordance with Eq.~\eqref{eq:TailSimplification} (compare also Fig.~\ref{fig:Tail}).}
\label{fig:TimeDomain}
\end{figure}

The rescaled $l=2,m=2$ signal as measured at future null infinity is plotted in Fig.~\ref{fig:TimeDomain} for $\epsilon\equiv 1-a=\{10^{-2},10^{-3},10^{-4},10^{-6}\}$. The transition between exponential and polynomial ringdown, also shown in Fig.~\ref{fig:Tail}, and its dependence on $\epsilon$ is clearly visible. In particular, we see that the time during which polynomial ringing is observed increases with decreasing $\epsilon$, as indicated by Eq.~\eqref{eq:TailSimplification}.

\begin{figure}
\includegraphics[width=\columnwidth]{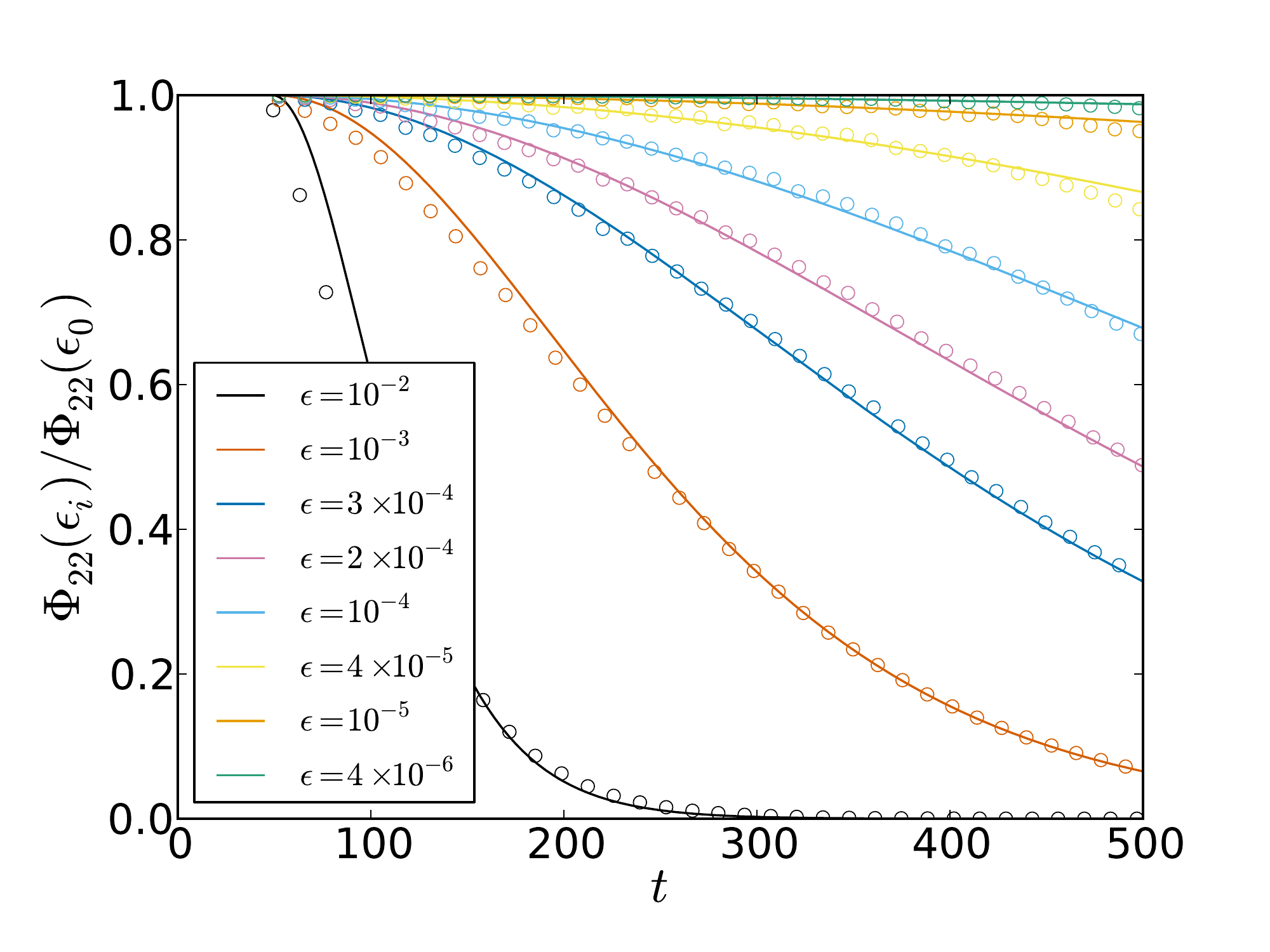}
\caption{Comparison of analytic formula and numerical calculation after the peak. The dots are the numerical maxima in the  previous figure. The solid lines are computed from the analytic formula Eq.~\eqref{eq:EpsRatio}.}
\label{fig:Epsilons}
\end{figure}

To test our analytic expressions quantitatively, we compute the ratio of the signals for two values of $\epsilon$. The leading-order behavior of the real amplitude for the $l=2,m=2$ mode from Eq.~\eqref{eq:asympt} reads
\be \label{eq:EpsRatio} \Phi_{22}(t;\epsilon) \approx  \sqrt{\epsilon} \frac{e^{-\sqrt{\epsilon/8} t}}{1-e^{-\sqrt{\epsilon/2} t}} \ee
To cancel out the effect of neglected constants in this formula we plot the ratio $\Phi_{22}(\epsilon_i)/\Phi_{22}(\epsilon_0)$ in Fig.~\ref{fig:Epsilons} for various listed values of $\epsilon_i$, and $\epsilon_0=10^{-6}$. The numerical values correspond to the local maxima of the oscillations, and are depicted by empty circles. The analytical approximation given by the ratios of Eq.~\eqref{eq:EpsRatio} is depicted by solid lines. The remarkable agreement indicates that our approximation is valid for a wide range of large angular momenta.

\section{Conclusions}
\label{sec:Conclusions}

In this paper, we systematically analyzed the QNM spectra of NEK black holes. For ZDMs, we applied the matched asymptotic expansion technique to derive the Hod formula and its error term, and showed the existence of ZDMs for all $m \ge 0$ modes; for DMs, we used the WKB method to approximate their frequencies, and gave a physical explanation to associate the existence of these modes with the presence of a potential peak outside the horizon (see Appendix~\ref{ap:DetTransform}). 
In the case that DMs and ZDMs coexist, as one decreases the black hole angular momentum, these two sets of modes merge at some finite $a$, and eventually become a single branch of modes. In order to understand this process better, we introduced a technique to transform the radial Teukolsky equation from a scattering problem to a bound-state problem. 
After this transformation, the new wavefunctions of the modes generally have finite support in the $r$ direction, which is a useful property to spatially distinguish ZDMs and DMs.   

An application of the NEK QNMs is to study the QNM response of a NEK black hole to perturbations. By applying the Hod frequency formula, we showed that the collective behavior of many superimposed overtones results in a power-law decay in the early part of the signal, which decays exponentially later on, and eventually becomes a polynomial tail due to scattering off the curvature of the radial Teukolsky potential. We also compared the approximate analytical formulae with numerical results obtained using the spectral Einstein code (SpEC) to evolve scalar waves of the Kerr background, showing close agreement for a wide range of angular momenta with $\epsilon \ll 1$.

It is interesting to conjecture a mode bifurcation for other nearly extremal black holes. For example, near-extremal Reissner-Nordstr\"{o}m black holes show no mode-branching \cite{Onozawa:1995vu, Andersson:1996xw, Konoplya}. For generic Kerr-Newman black holes this is non-trivial to verify, because perturbation fields with different spins are generally coupled with each other, and the coordinate dependence is not separable (but see \cite{Pani:2013ija} for recent progress in this direction). In addition, using the ``scattering/bound-state" transformation, in Fig.~\ref{fig:wavefunctionnearhorizon} we showed the ZDM wavefunction for the first $5$ overtones in the critical rotating limit $a=1$. It will be instructive to investigate both the ZDM and DM wavefunctions near the bifurcation point, and to see how the two sets of modes parametrically bifurcate by tuning $a$.

\acknowledgments
We thank David Nichols and Zhongyang Zhang for discussions during the early stages of this work, and Sam Dolan for advice on the WKB method,. This research is funded by NSF Grants PHY-1068881 and PHY-1005655, CAREER Grants PHY-0956189 and PHY-1055103, NASA Grant No. NNX09AF97G, the Sherman Fairchild Foundation, the Brandon Foundation, and the David and Barbara Groce Startup Fund at Caltech.

\appendix

\section{Wave amplitudes for NEK perturbations}
\label{ap:NEKCalcs}

We collect some of the derivations and results associated with the matching of the radial functions in the NEK spacetime as derived in Sec.~\ref{sec:NearlyExtreme}. We also give an expression for the amplitude of the QNM response discussed in Sec.~\ref{sec:Decay}. The coefficients $A$ and $B$ of the asymptotic $(x = r - M \to \infty)$ NEK perturbations given by matching to the interior solution and boundary conditions are
\begin{align}
A & = \frac{ \left( \sqrt { 8 \epsilon} \right)^{1/2 - i \delta} \Gamma(1 - i \sqrt{2} \tilde \omega) \, \Gamma(2 i \delta)}{\Gamma( 1/2 + i m + i \delta - i \sqrt{2} \tilde \omega) \, \Gamma ( 1/2 - i m + i \delta)}\,,  \\
B & = A|_{\delta \to - \delta} .
\end{align}
By considering the asymptotic solution as $r \to \infty$ of Eq.~\eqref{eqconfsln}, we can use our expressions for $A, \, B$, and the conversion between the radial functions $R$ and $u$ to find the ingoing and outgoing wave amplitudes. For scalar excitations, these are defined by $u \to A^{\rm out} e^{i \omega r_*} + A^{\rm in} e^{-i \omega r_*} $ as $r\to \infty$, when the amplitude of waves entering the black hole $A^{\rm hole}$ is set to unity to normalize the amplitudes. They are
\begin{widetext}
\begin{align}
\label{eq:Ain}
A^{\rm in} & = \left( \frac{e^{i \pi}}{im}\right)^{1/2 + im + i \delta} e^{(- i \delta/2) \ln 8 \epsilon} ( 8 \epsilon)^{1/4} \frac{\Gamma[ 2 i \delta] \, \Gamma(1 + 2 i \delta) \, \Gamma(1 - i \sqrt{2} \tilde \omega)}{\Gamma^2(1/2 - im + i \delta) \, \Gamma(1/2 + im + i \delta - i \sqrt{2} \tilde \omega)} + ( \delta \to - \delta) \,, \\
A^{\rm out} & = (im)^{-1/2+im - i \delta} e^{(- i \delta/2) \ln 8 \epsilon} ( 8 \epsilon)^{1/4} \frac{\Gamma(2 i \delta) \, \Gamma(1+ 2 i \delta)\, \Gamma( 1 - i \sqrt{2} \tilde \omega)}{\Gamma(1/2 + i m + i \delta) \, \Gamma(1/2 - i m + i \delta) \, \Gamma( 1/2 + im + i \delta - i \sqrt{2} \tilde \omega)} + (\delta \to - \delta) \,.
\end{align}
\end{widetext}
The following simplifications occur when we evaluate at the ZDM frequencies:
\begin{align}
 \Gamma( 1/2 + im + i \delta - i \sqrt{2} \tilde \omega) & \to \Gamma ( -n + 2 i \delta) \,, \\
 \Gamma( 1/2 + im - i \delta - i \sqrt{2} \tilde \omega) & \to \Gamma ( -n - i \sqrt{2} \eta ) \notag \\ &\approx \frac{(-1)^n}{n!(- i \sqrt{2} \eta)} \,.
\end{align}
Now, in the second term in the expression for $A^{\rm out}$ (where $\delta \to - \delta$), there is a Gamma function in the denominator which is near its pole, $\Gamma(1/2 + im - i \delta - i \sqrt{2} \tilde \omega)$, and this means that the term is $O(\eta)$ and is negligible. Similarly, the second term in $A^{\rm in}$ would seem to be negligible, but it actually dominates the derivative of $A^{\rm in}$.

In order to compute the excitation of the QNMs in Sec.~\ref{sec:Decay}, we need to compute $A^{\rm out}/ (d A^{\rm in}/ d\omega)$, evaluated at the roots of $A^{\rm in}$ (the QNM frequencies). By writing our terms as functions of $\tilde \omega$, we have expanded about the extremal frequency where amplitudes are evaluated, and in fact all of the factors of $m$ can be converted into factors of $2 \omega$. In order to avoid the unnecessary complication of converting back these factors, we take $d / d \omega = (1/ \sqrt{\epsilon}) d/ d \tilde \omega =  (1/ \sqrt{\epsilon}) d/ d \eta$, and insert the expression~\eqref{eq:PoleSln} to eliminate $\tilde \omega$ in favor of $\eta$ in the expression for $A^{in}$. Applying the $\eta$ derivative to $A^{\rm in}$, only the application of the derivative to the functions of $\eta$ in the denominators gives nonzero terms at the QNM frequencies, and in fact only the second term survives; we have
\begin{align}
\frac{d A^{\rm in}}{d \eta} \propto \frac{\Gamma'(-n - i \sqrt{2} \eta)}{\Gamma^2(-n - i \sqrt{2} \eta)} & \to (-1)^{n+1} n!\,,
\end{align}
which cancels out the corresponding factor of $n!$ in $A^{\rm out}$. Putting it all together, we recover Eq.~\eqref{eq:ExFactors}, with the approximately $n$-independent term $C(m, \delta)$ given by
\begin{align}
C(m,\delta)  =& - \frac{i}{ \sqrt{2}}e^{2 i(m - \delta) \ln{|m|}} \frac{\Gamma^2(2 i \delta)}{\Gamma^2(-2 i \delta)} \notag \\ & \times
\frac{\Gamma^2( 1/2 - i m - i \delta)}{\Gamma(1/2 + i m + i \delta) \Gamma(1/2 - im + i \delta)} \,.
\end{align}
This term contributes to the overall amplitude and phase of the QNM ringing.

\section{Numerical Methods for Computing QNM Frequencies}
\label{ap:NumericalMethods}

To compute the QNM frequencies more accurately when $a \lesssim 1$, we use a
modified version of Leaver's continued-fraction algorithm~\cite{leaver}. 
In order to search for the QNMs of the NEK spacetime, we actually search for $\tilde \omega$ by substituting Eq.~\eqref{eq:tildeomega} into the standard expressions for the continued fraction, and then simplifying them algebraically.
The most significant difference between Leaver's and our methods arises in the
solution to the angular Teukolsky equation. When $s\neq 0$, we use the series expansion of Fackerell and Crossman \cite{Fackerell} (see 
also Appendix B of Fujita and Tagoshi \cite{Fujita}) to express the angular 
eigenvalue in terms of a continued fraction that depends upon the frequency
of the mode.
(Fackerell and Crossman expand the angular Teukolsky function in a series of
Jacobi polynomials, whereas Leaver finds his solution in terms of powers of
$1+\cos\theta$.) When $s= 0$, we use {\it Mathematica}'s built-in function for the eigenvalue of the spheroidal harmonic equation to find ${}_0 A_{lm}$.
For the radial Teukolsky function, we compute Leaver's expansion, but we 
specialize his expressions for nearly extremal angular momenta, $a=1-\epsilon$.
We then compute the continued-fraction solution for the frequency in terms
of the angular separation constant. 

To explicitly find QNMs, as we do e.g. for the $(2,2)$ modes of Fig.~\ref{fig:qnm22}, we use  {\it Mathematica}'s nonlinear root-finding algorithm. We can then solve the set of two 
continued-fraction equations to find the frequency and separation constant of
a mode. Because nonlinear root finding often requires an initial condition for the 
algorithm that is close to the actual solution, we find that two different strategies are useful in this root search. 
For high angular momenta, we find that seeding our root-finding algorithm with our analytic ZDM formula assists in finding the roots as we increase the angular momentum towards $a \to 1$.
At moderately high angular momenta for the ZDMs, and for the DMs, we find that iteratively using the previous frequency value to seed the root finding as we raise (or lower) the angular momenta helps to track a given root. We also utilize inversion of the continued fraction~\cite{leaver} to find higher overtones, especially for lower angular momenta where the overtones have a well-defined ordering. 

Despite these techniques, we find it difficult to track a single root when many QNM values are close together, such as for very high angular momenta where the ZDMs begin to cluster.

\section{Criteria for the phase boundary of electromagnetic and gravitational modes}
\label{ap:DetTransform}

For electromagnetic and gravitational perturbations, the radial Teukolsky 
potential for extremal Kerr is a complex function. 
To obtain a real potential and a well-defined peak, we apply the 
transformations described in \cite{detweiler2}. 
Before the transformation, the radial Teukolsky equation is
\begin{align}
\Delta^{-s}&\frac{d}{dr}\left ( \Delta^{s+1}\frac{d}{dr}{}_s R\right ) \nonumber \\
& +\frac{K^2+i s \Delta' K-\Delta(2 i s K'+{}_s \lambda_{lm})}{\Delta}{}_s R =0\,,
\end{align}
where ${}_s R$ and $u$ are related by $u=\Delta^{s/2}(r^2+a^2)^{1/2} \,{}_s R$, and the primes indicate derivatives with respect to $r$ throughout this section. A new field variable $X$ can be defined by
\be
X=\Delta^{s/2}(r^2+a^2)^{1/2}\left [\alpha(r){}_s R+\beta(r)\Delta^{s+1}\frac{d {}_s R}{d r} \right ]\,.
\ee
The functions $\alpha(r)$ and $\beta(r)$ can be chosen such that the master equation satisfied by $X$ has a real valued potential. For electromagnetic perturbations ($s=-1$) the corresponding transformation is
\be
\alpha=\frac{\tilde{a}\Delta+1}{\kappa^{1/2}[{\rm Re}(\tilde{a}\Delta)+1]^{1/2}}\,, 
\quad 
\beta=\frac{\tilde{b} \Delta}{\kappa^{1/2}[{\rm Re}(\tilde{a}\Delta)+1]^{1/2}}\,,
\ee
where
\begin{align}
&\tilde a=[4K^2+2\Delta(i K'-\lambda)]/\Delta^2\kappa,\, \quad 
\tilde b=-4 i K/\Delta \kappa\,, \nonumber \\
&\kappa=(4\lambda^2-16a^2\omega^2+16a\omega m)^{1/2}\,,
\end{align}
and where we have dropped the harmonic indices on $_s \lambda_{lm}$. The potential is
\begin{align}
V_r &= \frac{-K^2+\lambda\Delta}{(r^2+a^2)^2}-\frac{\Delta r(\Delta r+4 M a^2)}{(r^2+a^2)^4} \nonumber \\
&+\frac{\Delta[\Delta(10r^2+2\nu^2)-(r^2+\nu^2)(11r^2-10r M+\nu^2)]}{(r^2+a^2)^2[(r^2+\nu^2)^2+\eta\Delta]} \nonumber \\
&+\frac{12\Delta r(r^2+\nu^2)^2[\Delta r-(r^2+\nu^2)(r-M)]}{(r^2+a^2)^2[(r^2+\nu^2)^2+\eta\Delta]^2} \nonumber \\
&-\frac{\Delta(r-M)^2\eta[2(r^2+\nu^2)^2-\eta\Delta]}{(r^2+a^2)^2[(r^2+\nu^2)^2+\eta\Delta]^2},
\end{align}
where we have restored factors of $M$ and defined
\be
\nu^2=a^2-a m/\omega,\, \quad \eta=(\kappa-2\lambda)/(4\omega^2).
\ee
As with our analysis of scalar modes, we take the limits $a\to 1$ and 
$\omega\to m/2$ and check whether there is a peak in the potential outside the 
horizon. 
After some calculation, we arrive at the condition for the existence of such a 
peak:
\be\label{eqeleccri}
{}_{-1}A_{lm}>\frac{7}{4}m^2\,.
\ee

For gravitational perturbations ($s=-2$) the transformation involves the functions
\be
\alpha=\frac{2(a_1\Delta^2i1a_2\Delta^2+|\kappa|)}{|\kappa|(a_1\Delta^2+|\kappa|)^{1/2}},\,
\beta=\frac{2ib_2\Delta^2}{|\kappa|(a_1\Delta^2+|\kappa|)^{1/2}}\,,
\ee
with
\begin{align}
a_1&=\frac{8K^4}{\Delta^4}+\frac{8K^2}{\Delta^3}\left ( \frac{M^2-a^2}{\Delta}-\lambda\right )
 \nonumber \\
&+\frac{4\omega K}{\Delta^3}(3r^2+2M r-5a^2)+\frac{12r^2\omega^2+\lambda(\lambda+2)}{\Delta^2}\,, 
\nonumber \\
a_2 &=\frac{-24\omega r K^2}{\Delta^3}-\frac{4\lambda(r-M)K}{\Delta}+4\omega r \lambda+12\omega M\,, \nonumber \\
b_2 &= -\frac{8 K^3}{\Delta^2}-\frac{4 K}{\Delta}\left [\frac{2(M^2-a^2)}{\Delta}-\lambda\right ]-\frac{8\omega}{\Delta}(Mr-a^2)\,,
\end{align}
and
\begin{align}
\kappa &=[\lambda^2(\lambda+2)^2+144a^2\omega^2(m-a\omega)^2-a^2\omega^2(40\lambda^2-48\lambda) \nonumber \\
&+a\omega m(40\lambda^2+48\lambda)]^{\frac{1}{2}}+12i \omega M\,,
\end{align}
and the new potential term is
\begin{align}
V_r=\frac{-K^2+\Delta\lambda}{(r^2+a^2)^2}+\frac{\Delta(b_2p'\Delta)'}{(r^2+a^2)^2b_2p}+G^2+\frac{d G}{d r_*}\,,
\end{align}
where
$p=(a_1\Delta^2+|\kappa|)^{-1/2}$.
It turns out this potential gives the following criterion for existence of a
peak outside the horizon:
\be\label{eqgravcri}
{}_{-2}A_{lm}>\frac{7}{4}m^2-2\,.
\ee

There is another transformation listed in \cite{detweiler2} which also gives a 
master equation with real-valued potential. 
After repeating the calculation above for the alternative transformation, it 
can be shown that Eq.~(\ref{eqgravcri}) remains valid for the new potential. 
Combining Eq.~(\ref{eqeleccri}), Eq.~(\ref{eqgravcri}) and the criterion for 
the scalar modes, the condition for generic spin of the perturbations is 
summarized by Eq.~(\ref{eqcriall}).

\section{The $s= 0$, $l=2$, $m=1$ QNMs}
\label{sec:ExtraMode}

\begin{figure*}[t!]
\includegraphics[width=0.45\textwidth]{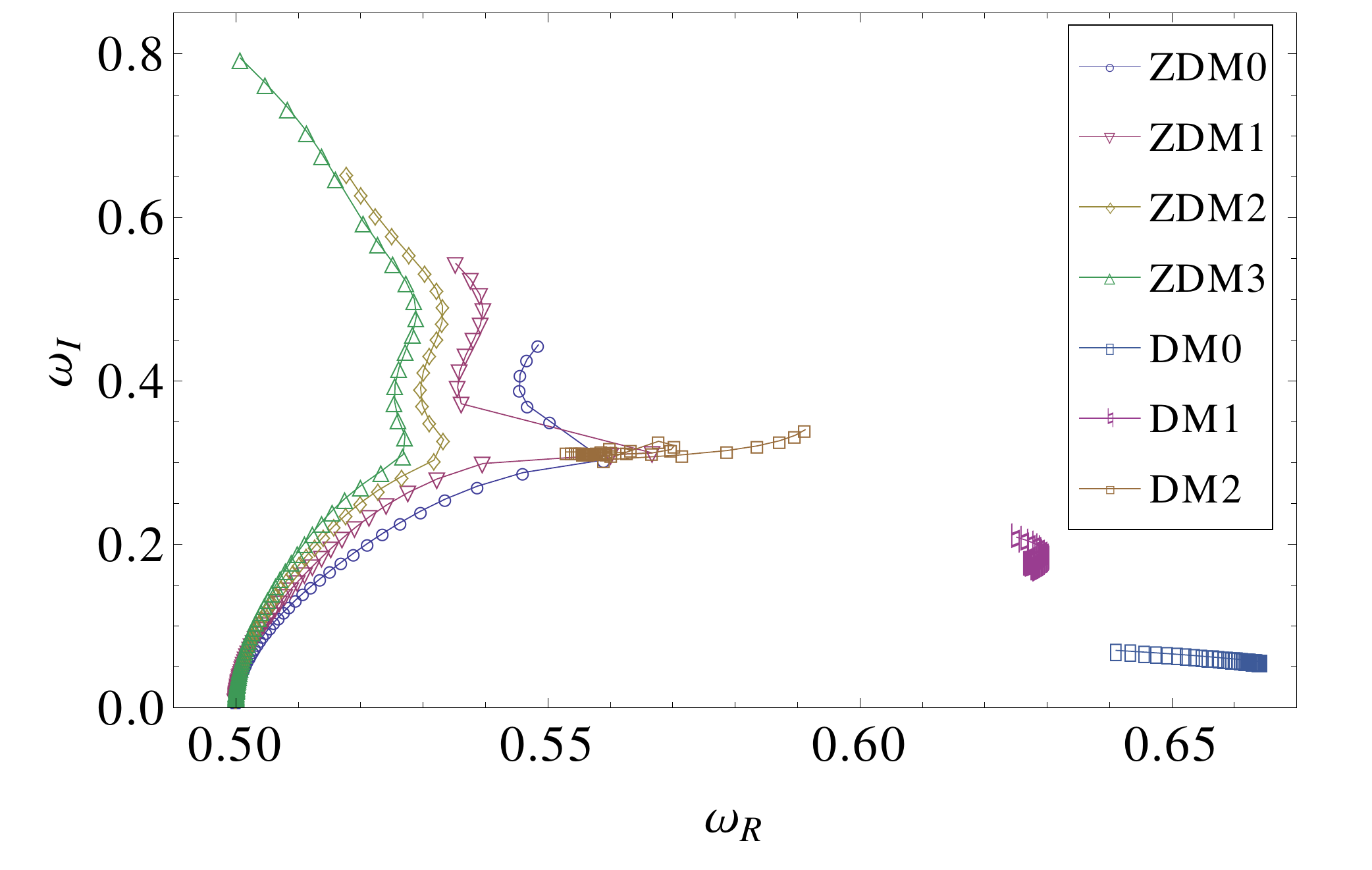}
\includegraphics[width=0.45\textwidth]{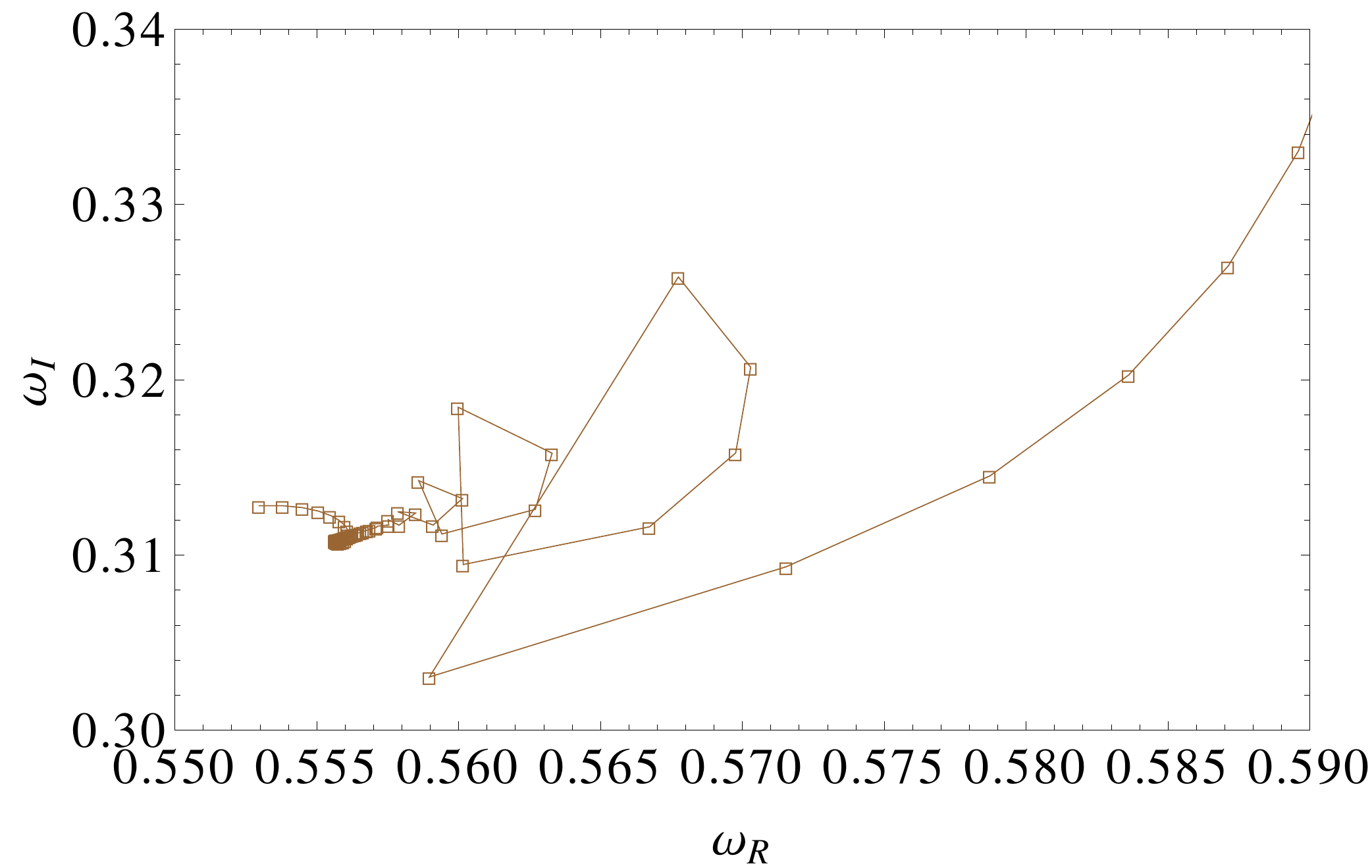}
\caption{Plots of QNM frequencies $\omega$ for the case $s = 0$, $(l,\,m) = (2,1)$, values as found using Leaver's method (with inversion). The first 7 overtones (at low angular momenta) are shown, which become for three DMs and the first four ZDMs as we increase $a$ from $0.950$ to $0.99999$ using logarithmically decreasing spacing. Left panel: All seven modes visualized. The first two ZDMs momentarily join the third DM, as discussed in the text. Right panel: The third DM executes spiraling motion, which we have not attempted to resolve; the first two sharp jumps occur because of interaction with the first two ZDMs.}
\label{fig:Modes21s0}
\end{figure*}

\begin{figure*}[t,b]
\includegraphics[width=1.0\textwidth]{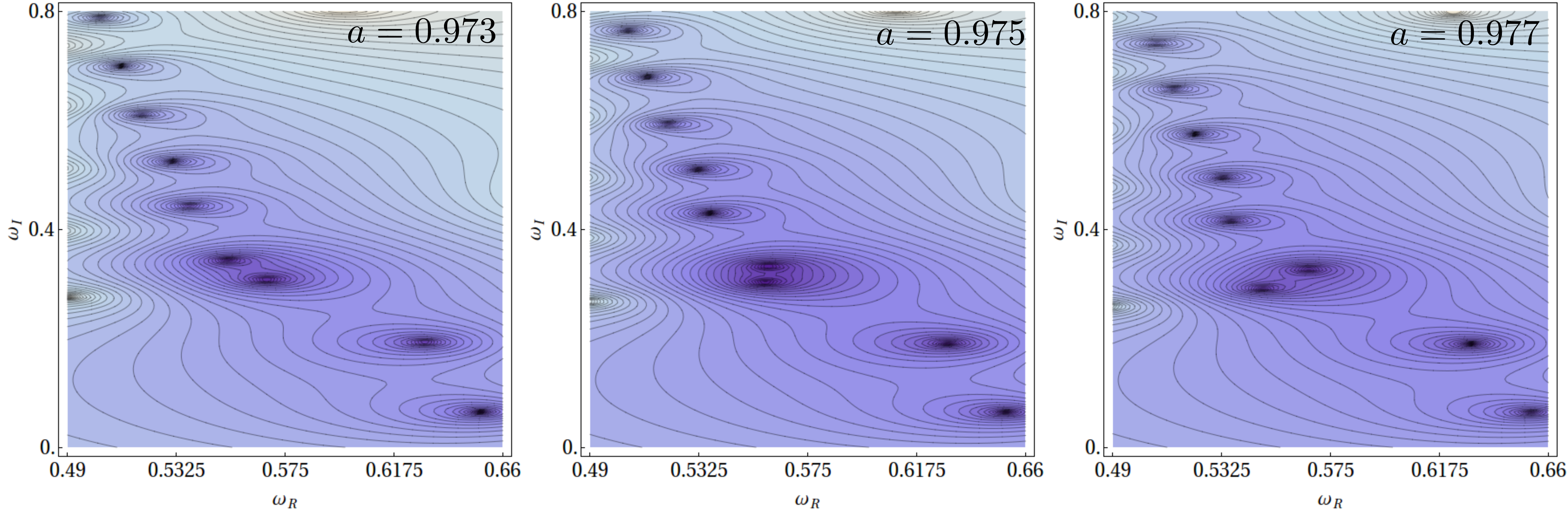}
\caption{The exchange of QNM identity for the $s=0$, $(2,1)$ mode, illustrated by continued fraction value as plotted in Fig.~\ref{fig:Modes21s0}. Top panel: At $a=0.973$, the third and fourth overtones approach each other. Middle panel: At $a= 0.975$ the fourth overtone lies just above the third. Bottom panel: As we raise the angular momentum to $a = 0.977$, we see the third overtone begins to move toward the real axis, becoming the first ZDM, while the fourth becomes the third DM.}
\label{fig:ModeCrossing}
\end{figure*}

In this Appendix, we discuss the results of our root search for $s =0$, $m=1$. We plot the trajectories of the first seven overtones as we increase the angular momentum in the left panel of Fig.~\ref{fig:Modes21s0}. 
Many of the features of this mode are shared with the $s = -2$, $(2,1)$ mode discussed in Sec~\ref{sec:L2GravModes}, but there are some new behaviors. 
In this case, there are three DMs rather than two, and the third DM executes a spiraling motion towards smaller $\omega_R$ with increasing angular momentum. Meanwhile, the next four overtones are ZDMs and move to steadily smaller values of $\omega_I$ with increasing angular momentum. The first two of these frequencies have rapid jumps in value as they near the third DM, as if strongly attracted to the continued fraction root there, before returning to a trajectory with steadily decreasing frequency and decay rate. Meanwhile, the right-hand panel of Fig.~\ref{fig:Modes21s0} shows that the third DM executes a spiraling trajectory that we find difficult to resolve, for the reasons discussed below. The final spectrum is completely separated into two branches, and the overtone $n$ QNMs at low angular momentum become the $n' = n-3$ ZDM overtones at high angular momentum.

We now examine more closely the behavior of the QNMs as the branches split. As the angular momentum approaches the value where the third QNM (third DM) and fourth QNM (first ZDM) overtones have nearly the same frequency, the frequencies approach each other, and for a value $a_s \approx 0.975$ they lie just above and below each other in the complex plane. Beyond the splitting angular momentum $a_s$, the modes actually switch roles, with the fourth overtone moving towards larger $\omega_R$ and becoming the third DM, while the third overtone begins to move to lower decay rate. 

The root exchange is illustrated in Fig.~\ref{fig:ModeCrossing} using contours of constant logarithm of the continued fraction, as in Fig.~\ref{fig:Modes21s0}.
Our root finding routine does not resolve this behavior, and instead switches which root it follows just as this splitting occurs. This account for the first large jump in the trajectory of the third DM seen in the right panel of Fig.~\ref{fig:Modes21s0}, which occurs at about $a_s = 0.975$, when the frequencies exchange. A similar investigation shows that when the fifth overtone nears what has become the third DM, the two modes approach closely and then scatter away from each other without switching roles; this occurs between $a = 0.98660$ and $a = 0.98665$. This is also when the spiral in the right panel of Fig.~\ref{fig:Modes21s0} experiences its second sharp increase in $\omega_I$, at $a = 0.986$. The origin of these odd behaviors is unclear.

\bibliography{refs}

\end{document}